\documentclass[usenatbib]{mn2e}

\usepackage{graphicx}
\usepackage{xspace}
\usepackage{txfonts} 
\usepackage{url}

\def\lenstool{\textsc{lenstool}\xspace}

\newcommand{\rcrc}{r_{cut}/r_{core}}

\begin{document} 

\title[Multi-scale cluster lens mass mapping]{Multi-scale cluster lens
mass mapping\\ I. Strong Lensing modelling}

\author[E.~Jullo \& J.-P.~Kneib]
{E.~Jullo$^1$\thanks{E-mail: eric.jullo@oamp.fr}, J.-P.~Kneib$^1$ \\
$^1$ Laboratoire d'Astrophysique de Marseille, CNRS-Universit\'e de
Provence, 38 rue F. Joliot-Curie,  13013 Marseille, FRANCE 
} 

\maketitle

\begin{abstract}
We propose a novel technique to refine the modelling of galaxy
clusters mass distribution using gravitational lensing. The idea is to
combine the strengths of both ``parametric'' and ``non-parametric''
methods to improve the quality of the fit. We develop a multi-scale
model that allows sharper contrast in regions of higher density where
the number of constraints is generally higher.  Our model consists of
(i) a multi-scale grid of radial basis functions with physically
motivated profiles and (ii) a list of galaxy-scale potentials at the
location of the cluster member galaxies.  This arrangement of
potentials of different sizes allows to reach a high resolution for
the model with a minimum number of parameters. We apply our model to
the well studied cluster Abell~1689. We estimate the quality of our
mass reconstruction with a Bayesian MCMC sampler.  For a selected
subset of multiple images, we manage to halve the errors between the
positions of predicted and observed images compared to previous
studies. This owes to the flexibility of multi-scale models at
intermediate scale between cluster and galaxy scale. The software
developed for this paper is part of the public lenstool package which
can be found at www.oamp.fr/cosmology/lenstool. 

\end{abstract}


\section{Introduction}

Non-baryonic dark matter is today commonly accepted as a predominant
contributor to the matter density of our Universe. Dark Matter is indeed required to explain the velocity distribution of stars and gas in galaxies \citep[e.g][]{salucci2001,deblok2003} or the velocity dispersion of the galaxies in galaxy clusters \citep[e.g.][]{czoske2002} but also to reproduce the large scale galaxy distribution
\citep[e.g.][]{guzzo2008,seljak2005} and the cosmological microwave
background (CMB) fluctuations \citep[e.g. the review by ][]{hu2002}.
A particularly striking demonstration of the need of Dark Matter is
the direct detection of Dark Matter in the ``Bullet Cluster''
\citep{clowe2006,bradac2006} a cluster merger of 2 massive clusters
where the X-ray gas and DM are spatially segregated due to the
difference in nature of baryons and DM.  However, beyond this general
agreement about the need of dark matter, very little is known about
its nature.  The increasing amount of observational evidences steadily
rules out possibilities. For instance, the CMB seems to rule out the
warm dark matter \citep{spergel2003} and the measured extension of the
truncation radius of galaxies in clusters seems to rule out fluid-like,
strongly interacting dark matter at 5$\sigma$ \citep{natarajan2007a}. 
On another
aspect, numerical simulations show that the collapsed non collisional
dark matter forms NFW density profiles \citep*{navarro1997}, with a
central density peak.  In contrast, if self-interacting particles are
considered, this peak can be replaced by a profile with flat core
\citep{spergel2000}. On the observational side, measuring the slope of
the dark matter density profile is still a hot topic either at the
galaxy scale, with discrepancies in the rotation curves of stars
\citep{gentile2007,valenzuela2007}, or at the cluster scale, with
discrepancies related to the slope of density profiles determined
by gravitational lensing \citep{meneghetti2007,sand2008,limousin2008}.
Finally, numerical simulations with self-interacting particles predict less small scale
halos than simulations with non collisional particles. For a long time, the
\emph{missing satellite problem} in the Local Group was considered as
an evidence in favor of the self-interacting particles hypothesis.
However, \citet{simon2007} have shown that this problem could merely
be an observational issue. On the cluster scale,
\citet{natarajan2007a} has shown that the mass function of
substructures was in agreement with simulations with non-collisional
particles, at least
for a few strong lensing clusters.  Estimating the mass
distribution of cosmological objects with great accuracy is therefore
a unique way to unveil the nature of dark matter.

Since the early 90's, gravitational lensing has appeared as a robust
tool to model the mass distribution of cosmological objects alike
galaxies, galaxy clusters and large scale structures
\citep{gavazzi2007,limousin2007b,massey2007nat,fu2008}. With deep
HST/ACS observations of massive clusters of galaxies, a large number
of multiple images have been uncovered. In particular in the case of
Abell 1689, \citet{broadhurst2005} were the first to identify more
than 100 multiple images part of more than 30 multiple images
systems. However, in such massive clusters, strong lensing 
modelling has been unable to reproduce the numerous systems of multiple images with less than typically 1'' residual RMS. In Abell~1689,
\citet{limousin2007b} obtain an RMS of 2.87'' for 34 systems of
multiple images, \citet{halkola2006} report an RMS of 2.7'' and
\citet{broadhurst2005} an RMS of 3.2''. In Abell~1703,
\citet{limousin2008} obtain an RMS of 1.45'' with 13 systems of
multiple images, and in Abell~2218, \citet{eliasdottir2007b} obtain an
RMS of 1.49'' with 8 systems.  The physical origin of this
systematical error is not yet fully understood. Do we miss invisible
small scale subhaloes in our models, or are we badly reconstructing
the large scale mass distribution?

These large residual errors are likely highlighting a lack of resolution
and/or flexibility in the lensing mass models. Indeed, mass models 
traditionally consist of an analytical density profile centered
with respect to the light distribution, and fitted to the positions of
the multiple images. In addition, \citet{kneib1996} have shown that
the complementary modelling of galaxy-scale halos hosting bright
cluster member galaxies significantly improves the fit \citep[see][for
a throughout description of the analytical modelling of
clusters]{jullo2007}. In contrast
to traditional ``parametric'' modelling of galaxy clusters,
partisans of ``non-parametric'' models claim that their methods may
allow a perfect fit; however at the expenses of sometimes unphysical
solutions.  In the case of ``non-parametric'' method, the mass
distribution is generally tessellated into a regular grid of small
elements of mass, called pixels \citep{saha1997,diego2005a}.
Alternatively, \citet{bradac2005} prefer tessellating the
gravitational potential because its derivatives directly yield the
surface density and other important lensing quantities.  
Point-like pixels can also be replaced by radial basis functions
(RBF). RBF are real-valued functions with radial symmetry.  Several
density profiles for the RBF have been tested so far. \citet{liesenborgs2007} use
Plummer profiles, and \citet{diego2007} use RBF with Gaussian
profiles. They also study the properties of Power law and Isothermal
profiles as well as Legendre and Hermite polynomials. They advise to
use compact-like profiles such as the Gaussian or the Power law
profiles, since too extended profiles produce a constant sheet excess
in the resulting surface mass density.  Finally, instead of using a
regular grid, \citet{coe2008} and \citet{deb2008} use the actual
distribution of images as an irregular grid. Then, they either place
RBF on this grid or directly estimate the derivatives of the potential
at the images location. 

Whatever their implementation, the multiple images reproduction is
generally greatly improved with respect to traditional ``parametric'' modelling.
However, the qualification of these models is still a matter of
debate. Indeed, because of their large number of free parameters with
respect to the number of constraints, many different models can
perfectly fit the data. In order to identify the best physically
motivated model and
eventually learn something on dark matter distribution in galaxy
clusters, external criteria (e.g. mass positivity) or regularization
terms (e.g. to avoid unwanted high spatial frequencies) are required.
In addition, galaxy mass scales are never taken into account, although
traditional modelling have demonstrated that they effectively affect
multiple images positions. This final step makes such
``non-parametric'' models a little uncertain.  Nonetheless,
``non-parametric'' models are useful because their large flexibility
allows the exploration of the mass distribution regardless of any {\it
a priori}.  For example, these ``non-parametric'' methods are
efficient to reveal complex mass distribution such as found in the
``Bullet Cluster''.

In this article, we study the properties of a model made of a
multi-scale grid of RBF and a sample of analytically defined
galaxy-scale DM halos. We analyse how this model compares to a
traditional ``parametric'' model. We apply our analysis to the galaxy
cluster Abell~1689 for its large amount of systems of multiple images.
In section 1, we present the analytic definition of our RBF.  In
section 2, we evaluate the ability of our multi-scale grid model in
reproducing a simple NFW profile. In section 3, we use the mass model
of \citet{limousin2007b} as an input to build a multi-scale grid model
of the galaxy cluster Abell~1689. In section 4, we fit this model to a
subset of multiple images, and compare the produced mass map,
deviation angle map and shear map to the ones obtained with a
traditional model. In addition, we perform an over-fitting check. To
do so, we assume that if a model optimised with a subset
of an image catalogue can accurately predict the rest of it, it does
not over-fit the data. Therefore, we compare the RMS between predicted
and observed images for the part of the image catalogue not used as
constraints.  Finally in section 5, we study how different values of
parameters related to the grid building affect the quality of the fit,
the density profile of the cluster, and the estimated properties of
the galaxy-scale halos.  When necessary, we use the flat $\Lambda$CDM
concordance cosmology with $\Omega_m=0.3$, and $H_{0}$=74 km/s/Mpc. At
the redshift of Abell~1689 $z=0.184$, an angular scale of 1 arcsec
corresponds to 2.992 kpc. 

\section{Lensing equations}

The lens equation~:

\begin{equation}
	\label{eq:lens}
	\beta(\theta)=\theta - \alpha(\theta)\;,
\end{equation}

\noindent defines the transformation between the image position
$\theta$ and the source position $\beta$. $\alpha(\theta)$ is the
deflection angle due to the lens \citep[e.g.][]{schneider1992}. 

The amplification $\mu$ of an image located in $\theta$ is inversely
proportional to the determinant of the amplification matrix $A$

\begin{equation}
	\mu(\theta) = \frac{1}{|\det(A)|}\;,
\end{equation}

\noindent where the amplification matrix $A$ is the derivative of the
lens equation at the image location
 
\begin{equation}
	A = \frac{\partial \beta}{\partial \theta} = \left[ \begin{array}{cc}
		1-\kappa+\gamma & 0 \\
		0 & 1-\kappa-\gamma 
	\end{array} \right]
\end{equation}

\noindent here expressed in the amplification basis. $\kappa$ is the
convergence and $\gamma$ is the shear.

Through the Fermat Principle, it is possible to demonstrate that the
deflection angle $\alpha(\theta)$ is proportional to the gradient of the
two-dimensional Newtonian potential \citep{blandford1986}

\begin{equation}
	\mathbf{\alpha(\theta)} = \frac{2}{c^2}
	\frac{D_{LS}}{D_{OS}} \nabla \phi(\theta)\;,
\end{equation}

\noindent which in turn is related to the surface density $\Sigma$ and
the convergence $\kappa$ through the Poisson relation in 2D

\begin{equation}
	\frac{2}{c^2} \frac{D_{LS}}{D_{OS}} \nabla^2 \phi(\theta) =
	\frac{2}{c^2} \frac{D_{OL} D_{LS}}{D_{OS}} 4 \pi G \Sigma =
	2 \frac{\Sigma}{\Sigma_{crit}} = 2 \kappa
\end{equation}

$\Sigma_{crit}$ is the critical density above which strong lensing is
possible. $D_{OL}$, $D_{LS}$ and $D_{OS}$ are cosmological angular
distances between the observer $O$, the lens $L$ and the source $S$. 

Deflection angles are additive quantities. For instance, if in a cluster we consider
$N$ clumps of mass located in $\theta_i$, then each of them independently
deflects a light beam crossing the cluster by an angle $\alpha_i$.
The total deflection angle computed at an observed image position
$\theta$ is then

\begin{equation} 
	\mathbf{\alpha(\theta)} = \sum_{i=1}^N
	\mathbf{\alpha_i(|\theta-\theta_i|)}\;.
\end{equation}

Let us now define the RBF used to build the multi-scale grid. Its
density profile is given by the analytical expression of the Truncated Isothermal Mass Distribution
(TIMD), which is the circular version of the
truncated Pseudo Isothermal Elliptical Mass Distribution (PIEMD)
\citep{kassiola1993,kneib1996,limousin2005,eliasdottir2007b}. 
The analytical expression of its surface density is  

\begin{equation}
	\label{eq:sdens}
	\Sigma(R)  =  \sigma_0^2 f(R,r_{core},r_{cut}) 
\end{equation}

\noindent with

\begin{equation}
	f(R,r_{core},r_{cut}) =  \frac{1}{2G}
		\frac{r_{cut}^2}{r_{cut}^2 - r_{core}^2}
		\left( \frac{1}{\sqrt{r_{core}^2 + R^2}} -
	\frac{1}{\sqrt{r_{cut}^2+R^2}}\right)\;.
\end{equation}

$f$ defines the profile, and $\sigma_0^2$ defines the weight of the
RBF. Note that this profile is characterised by two changes in its
slope marked by the $r_{core}$ and $r_{cut}$ radius \citep[see Fig.~1
in][]{jullo2007}. Within $r_{core}$, the density is roughly constant,
between $r_{core}$ and $r_{cut}$, it is isothermal $\Sigma \propto
r^{-1}$, and beyond $r_{cut}$ it falls as $\Sigma \propto r^{-3}$.
Compared to the Gaussian RBF profile used by \citet{diego2007} or the
Plummer RBF profile used by \citet{liesenborgs2007}, the TIMD profile
has a shallower slope in the centre but falls in a steeper manner after $r_{cut}$,
thus preventing from the mass sheet excess noted by \citet{diego2007}
with the pure isothermal profile.  In addition, the TIMD profile is
physically motivated. Its total mass is finite as well as its central
density. In this respect, this profile is more physical than the
notorious NFW potential \citep{navarro1997} which fits non collisional
dark matter numerical simulations but has an infinite central density
and an infinite total mass \citep[see][in which TIMD and NFW
potentials are compared]{limousin2005}.  Finally, thanks to its
flat core, the TIMD potential can produce extended flat regions, in
particular in the centre of clusters if necessary.

\section{The multi-scale grid}
\subsection{Definition and motivation}

\begin{figure}
	\centering
		\includegraphics[width=0.5\linewidth]{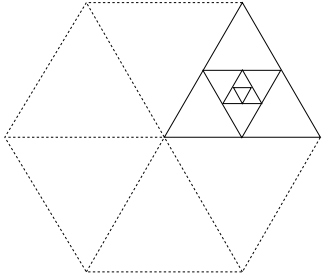}

	\caption{\label{fig:hexagon}
	Adopted hexagonal geometry for the multi-scale grid and
	recursive splitting into equilateral triangles.}

\end{figure}

In this section, we detail how we build the multi-scale model, and
demonstrate its capabilities at reproducing a singular NFW mass
profile. For the moment, we do not include lensing constraints.

As proposed by \cite{diego2005a}, we create a coarse multi-scale grid from a
pixelated input mass map and recursively refine it in the densest regions.
Doing so, the huge range in mass observed in galaxy clusters is
efficiently sampled with a minimum number of grid pixels. In contrast
to many previous works, we do not use a squared grid but an hexagonal
grid, on the ground that it better fits the generally rounded shape of
galaxy clusters.  With such a geometry, it is straightforward to
generate a triangular mesh, which is the best way to pack a set of
radial basis functions.

\begin{table}
	\centering
\scriptsize
	\caption{\label{tab:nbtri}
	\textsc{Characteristics of hexagonal grids per level of splitting}}
	\begin{tabular}{cccc}
		\hline
		Level & Nb. of triangles & Nb. of nodes &
		\parbox{1.7cm}{\centering Size of a triangle} \\
		\hline
		0  & 6  & 7    & 1 \\
		1  & 24  & 19  &  1/2\\
		2  & 96  & 61  & 1/4 \\
		3  & 384 & 217 &  1/8\\
		4  & 1536 & 817 & 1/16\\
		5  & 6144 & 3169  & 1/32\\
		6  & 24576 & 12481  & 1/64\\
		\hline
	\end{tabular}
\end{table}

In practice, we start by bounding the field of interest with an
hexagon centred on the cluster centre. We split it into 6 equilateral
triangles as shown in Fig.~\ref{fig:hexagon}.  Then, we choose a
simple splitting criteria. We have tested several criteria~: total
mass, standard deviation or amplitude of surface density variations in
a triangle and density of constraints, but none of them worked as well
as the surface density threshold. Considering  for instance an input
mass map in a FITS image with pixels of 1 arcsec$^2$, a triangle on
this image is split into 4
sub-triangles if it contains a single pixel (i.e. a region of 1
arcsec$^2$) that exceeds a user defined surface density threshold.
For instance, in order to trigger strong lensing regions, the
threshold can be set equal to the critical density in
$M_\odot/arcsec^2$ at the cluster redshift.  Over-critical regions
will be split whereas sub-critical regions will not. In the extreme
case where the mass map is everywhere greater than the threshold, it
results into a regular grid in which the number of triangles increases
as $3\times 2^{2n+1}$ and the number of triangle summits, or grid
nodes, as $N=1+3\times(2^{2n}+2^n)$, where $n$ is the level of
recursive splitting. Table~\ref{tab:nbtri} summarises for some levels
of splitting the maximum number of triangles and nodes a grid can
contain. The level of splitting or equivalently the finest grid
resolution ($\sim 2^{-n}$) is set by the user. As stated above,
we have tested models where the grid is refined at the multiple images
location.  Similarly as in \citet{coe2008}, such models
lead to perfect fits to the data.  However, we note that they also
easily get biased by the chosen set of data.

Finally, RBF described by TIMD potentials are placed at
the grid node location $\theta_i$. Their core radius is set equal to the
size of the smallest nearby triangle and their cut radius is set
equal to
three times the core radius (this is discussed in Section
\ref{sec:discussion}). Their weight $\sigma_i^2$ are obtained by
inverting the following system of $N$ equations 

\begin{equation}
	\label{eq:model}
	\left[ \begin{array}{ccc}
		M_{11} & \cdots & M_{1N} \\
		\vdots & \ddots & \vdots \\
		M_{N1} & \cdots & M_{NN} \\
	\end{array} \right]
	\left[ \begin{array}{c}
		\sigma_1^2 \\ \vdots  \\ \sigma_N^2 \\ 
	\end{array} \right] =
	\left[ \begin{array}{c}
		S_1 \\ \vdots  \\ S_N \\ 
	\end{array} \right]   
\end{equation}
\noindent with
\begin{equation}
	M_{ij} =  f_j(|\theta_i - \theta_j|,r_{core_j},r_{cut_j})\;.
\end{equation}

$S_i$ is the surface density read from the input mass map at
the grid node location $\theta_i$. $M_{ij}$ is the value of a RBF
with $\sigma_j^2=1$, 
centered on the grid node location $\theta_j$, and computed at a radius
$R=|\theta_i-\theta_j|$. The product $M_{ij}\,\sigma_j^2$ gives the
contribution of this RBF to the surface density $S_i$
(see Eq.~\ref{eq:sdens}).

\subsection{Reproducing a NFW profile with a multi-scale model}

\begin{figure}
	\centering
	\includegraphics[width=\linewidth]{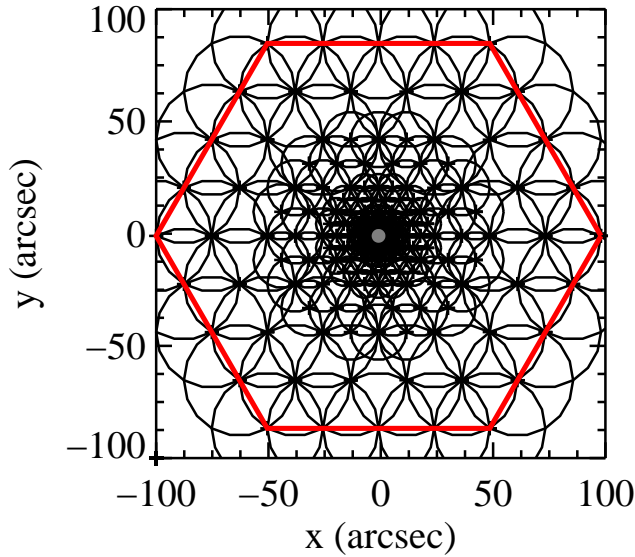} 

	\caption{\label{fig:nfw_split2D}
	Multi-scale grid made of 229 RBF, mapping an input mass
	distribution with an NFW profile. The radius of the circles
	corresponds to the core radius of the RBF, i.e. locally the
	grid resolution.  The central grey disk of radius $2\times
	R_6$ with $R_6 = 1.54''$ represents the central grid
	resolution beyond which mass map comparison is meaningless in
	Fig.~\ref{fig:nfw_split1D}.}

\end{figure}

\begin{figure}
	\centering
	\includegraphics[width=\linewidth]{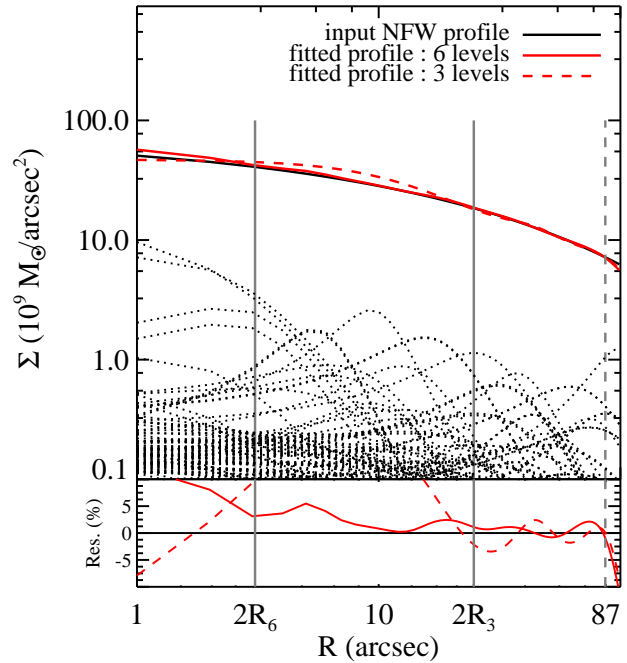}

	\caption{ \label{fig:nfw_split1D}
	Reproduction of the input NFW profile (in
	black) by a grid with 6 levels of splitting (solid red line)
	and 3 levels of splitting (dashed red line). The recovered
	profile with the 6 levels grid is the sum of the 229 RBF
	(dotted lines) shown in Fig.~\ref{fig:nfw_split2D}.  The two
	vertical lines at $2\,R_6$ and $2\,R_3$ mark the lower bounds
	of validity for the 6 levels grid and the 3 levels grid
	respectively.  The dashed vertical line marks the hexagon limit
	at $100\sqrt{3}/2$ arcsec.  Residuals 
	show that both models present similar levels of errors
	($<5\%$) within their respective domain of validity.  } 

\end{figure}

With enough resolution multi-scale models can reproduce all sorts of
input mass distributions, provided of course that no region has a
slope steeper than $\Sigma \propto r^{-3}$ (steepest slope of the TIMD
potential).  As an exercise, let us consider the input mass map
produced by a cluster  at a redshift $z = 0.184$ whose density profile 
follows a circular NFW profile with a concentration of 5 and a scale
radius of 150''. We compute a 200 x 200 arcsec$^2$ pixelated mass map
of this object (i.e. about an ACS field of view) to be used as input
mass map.  To build the multi-scale grid, we start with an hexagon
whose centre matches the centre of the NFW mass distribution and whose
radius is set to 100''.  We set the splitting threshold to $4.7\
10^{10} M\odot/arcsec^2$, and limit to $n=6$ the number of recursive
splitting. Fig.~\ref{fig:nfw_split2D} shows the produced multi-scale
grid.   The smallest triangle, i.e.  the smallest element of
resolution, is $R_6=\frac{100''}{2^6}=1.54''$ wide. The grid contains 229
nodes, to which we associate RBF with TIMD profile.  Their weight
$\sigma_i^2$ are obtained by inverting Eq.~\ref{eq:model}.
Note however, that we prefer an iterative method rather than a direct
matrix inversion.  Indeed, if we directly invert matrix $M$, we
obtain a perfect fit but the solution vector contains negative
$\sigma^2_i$, i.e. RBF with negative density profiles.  Although
\citet{liesenborgs2008a} allow some RBF to be negative, we are more
conservative and prevent any element from the $\sigma_i^2$ vector from
becoming negative. To do so, we minimise the following quantity

\begin{equation}
	Z = \sum_{j=1}^N \left( S_\mathrm{input}(\theta_j) -
	S_\mathrm{pred}(\theta_j) \right) ^2\;,
\end{equation}

\noindent where $S_\mathrm{input}$ and $S_\mathrm{pred}$ are the input
and predicted $S_i$ quantities of Eq.~\ref{eq:model}. With this
iterative inversion technique, we can force the RBF to have
positive weights $\sigma_i^2$, hence make sure that the overall
surface density is positive.  This way, we also avoid an additional
regularisation term (and possible related effects on the final
results) to control the sign of the surface density in the lensing
$\chi^2$ defined below in Eq.~\ref{eq:chi2}.

Fig.~\ref{fig:nfw_split1D} shows the reconstructed radial density profile.
Note how well the arrangement of RBF fits the input NFW density
profile. The residual is lower than 5\% on the meaningful domain from
twice $R_6$ to the hexagon inner limit. Note that a similar residual
is also achieved with 3 levels of splitting, though on the corresponding
meaningful domain.  Therefore, a large number of splitting is not
expressly justified unless the very centre of the mass distribution is
of particular interest regarding some strong lensing constraints (e.g.
radial arcs).

\section{Multi-scale model of Abell 1689}
\label{sec:multiscale}

\begin{figure}
	\centering
	\includegraphics[width=\linewidth]{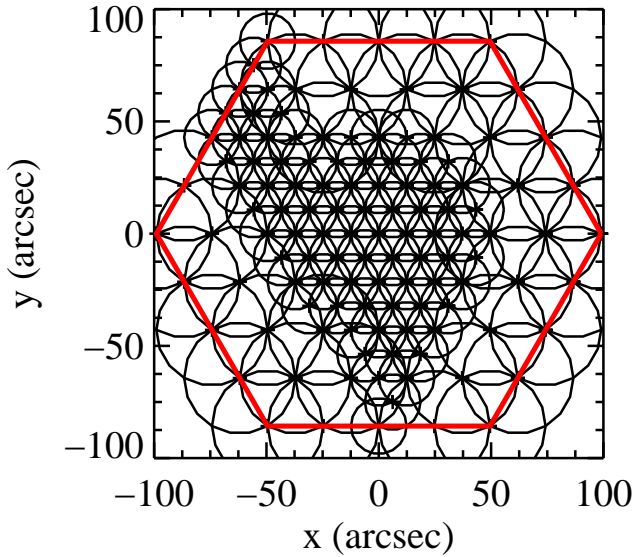}

	\caption{ \label{fig:grid_3levels}
	Multi-scale grid made of 120 RBF obtained with 3 levels of
	splitting, and based on the input mass map of Abell~1689
	\citep{limousin2007b}. }

\end{figure}

\begin{figure}
	\centering
	\includegraphics[width=\linewidth]{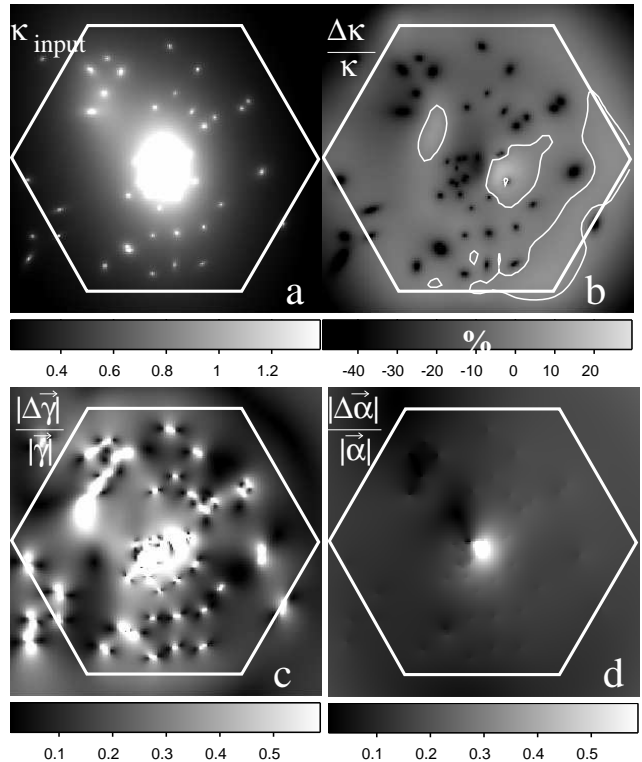}

	\caption{ \label{fig:res_wgals}
	{\bf (a)} Input convergence map of the galaxy cluster
	Abell~1689 for a source at redshift $z_s = 3$. The hexagon of
	radius 100'', bounds the region covered by the multi-scale
	grid. {\bf (b)} Relative error map
	$(\kappa_{grid}-\kappa_{input})/\kappa_{input}$ in percent. The
	white contours correspond to $\Delta\kappa=0$. On average, the multi-scale model
	underestimates the surface density. {\bf (c)} Relative shear
	error map $|\vec{\gamma}_{grid} - \vec{\gamma}_{input}| /
	|\vec{\gamma}_{input}|$. {\bf (d)} Relative
	defection angle error map $|\vec{\alpha}_{grid} -
	\vec{\alpha}_{input}| / |\vec{\alpha}_{input}|$. } 

\end{figure}

In this section, we continue to study the capabilities of multi-scale
models. However, in place of an NFW mass distribution, we build
the grid from a more realistic input mass map based on a simplified
version of the mass model of the galaxy cluster Abell~1689 reported in
\citet{limousin2007b}. In fact, our input model is made of 2 cluster-scale
and 60 galaxy-scale clumps of mass, instead of the 190 galaxy-scale
clumps used in Limousin et al. Only massive clumps producing
a deflection angle larger than 0.07'' (i.e. roughly the HST/ACS
imaging resolution) have been retained. All these clumps are described
by PIEMD potentials. The galaxy-scale clumps are associated to cluster
member galaxies. They follow scaling relations with similar properties as
the one reported in Limousin et al.  Fig.~\ref{fig:res_wgals}a shows
the convergence map produced by this model ( with $\Sigma_{crit} =
2\,10^{10}\,M_\odot/\mathrm{arcsec}^2$ calculated for a lens at
redshift $z_l = 0.184$ and a source at redshift infinity).

Following the same scenario as in the previous section, we start
building the multi-scale grid by bounding the field of interest with
an hexagon of radius 100'' centered on the image centre, also the
cluster centre. We set the splitting threshold to the critical surface
density stated above.  Doing so, we only oversample the central and
North-East strong lensing regions, within which $\kappa > 1$. Since the
particular distribution and shape of the observed multiple images
depend on the mass distribution in these regions, it is important to
give them as much flexibility as possible. In a first attempt, we stop
the levels of recursive splitting at 3. We obtain a grid of 120 RBF
shown in Fig.~\ref{fig:grid_3levels}. The size of the smallest
triangle hence our smallest element of resolution is $R_3 = 12.3''$.
In comparison, \citet{limousin2007b} estimate the core radius of the
cluster to 33'' which is more than twice $R_3$.  According to the
Shannon rule, we have therefore enough resolution to model it and
reproduce central systems of multiple images. The weights $\sigma_i^2$
of the RBF are computed by iterative inversion of Eq.~\ref{eq:model} as
detailed previously. 

Fig.~\ref{fig:res_wgals} shows a disagreement between the 
convergence map produced by this model and the input one, above all at the galaxy-scale clumps
location. Consequently, the multi-scale model underestimates by 9\%
the mass inside the hexagon. Beyond the hexagon limit, note the
increasing disagreement between the 2 models. This is a modelling
artefact of the multi-scale model due to
the difference in slope between the RBF, which falls as $\Sigma
\propto r^{-3}$, and the input profile slope, which falls as $\Sigma
\propto r^{-1}$.  Of course, such inaccuracies in the convergence map
induce serious discrepancies in terms of shear and deflection angle.
Fig.~\ref{fig:res_wgals}c shows large errors in the shear map at the
galaxy-scale clumps location, but smaller errors far from these
regions. In contrast, Fig.~\ref{fig:res_wgals}d shows that in the
deflection angle map, discrepancies are weaker, with only 10\% of
disagreement throughout.  Note that the discrepancy observed at the
very centre is due to the lack of resolution of the multi-scale
model. Therefore, we conclude that the lack of resolution of the
multi-scale model at the galaxy-scale clumps location severely affects
the lensing properties of the model. 

Of course, more levels of splitting could improve the recovery, as
demonstrated in the previous section with the NFW profile.  However
here, with 4 levels of splitting, the number of RBF raises to 318, and
with 6 levels, it amounts to 4287.  Optimising grids with so many
clumps is currently beyond our computational resources.

\subsection{Adding galaxy halos}

Instead of blindly increasing the level of splitting, we better build
a multi-scale model including both a multi-scale grid and galaxy-scale
clumps. The grid thus becomes a flexible mass component to fit the
cluster-scale distribution of mass, whereas galaxy-scale clumps fit
small-scale irregularities. This appears as a cheap solution to
our problem of resolution.

\section{Strong lensing constraints on multi-scale model}
\label{sec:constraints}

\subsection{Methodology}

In this section, we investigate the ability at constraining such
multi-scale models with strong lensing and compare the results in
terms of mass distribution and image prediction RMS, to a reference
analytical model. This comparison is performed on the observational data of the
galaxy cluster Abell~1689.  First, we describe the Bayesian MCMC
sampler we use to optimise the parameters.  Second, we present the
multi-scale and the reference models, their free parameters, assumed
priors, and the strong lensing constraints. Finally, we analyse the
results. 

\subsection{Bayesian MCMC sampler}

Given their large number free parameters and the few number of
constraints provided by strong-lensing data, ``non-parametric'' models
are usually under-constrained, hence the regularisation terms to
obtain smooth mass distribution
\citep{marshall2002,bradac2005,suyu2006}.  In contrast, we do not use
explicit regularisation term because our arrangement of RBF into a
regular grid, and the smooth and everywhere positive TIMD profile
describing them, constitute an intrinsic regularisation scheme. In
addition, equating the TIMD core radius of the RBF to the grid
resolution impedes strong discontinuities in the produced mass maps.
Finally, the overlap of nearby RBF correlates the model's parameters
and makes them dependent, thus reducing the effective number of free
parameters. The number of constraints may then become larger than the
number of effective parameters, and multi-scale models may not be
under-constrained (see Section~\ref{sec:discussion}).

In order to check our Bayesian MCMC sampler (based on the BayeSys library
\citep{skilling2004} implemented in \lenstool \footnote{publicly
available at \url{http://www.oamp.fr/cosmology/lenstool} }
\citep{kneib1993,jullo2007}), we generate a mock catalogue of 12
sources and 35 multiple images with the multi-scale model described
above. Then, we check that the mass distribution estimated from this
catalogue matches the input one. The $\chi^2$ is computed in the source
plane for simplicity and computation time. For each system $i$
containing $n_i$ multiple images, we define $\chi^2_i$ as

\begin{equation}
	\label{eq:chi2}
	\chi^2_i = \sum_{j=1}^{n_i} \left(
	\frac{\beta_j-\langle\beta\rangle}
	               {\mu_j^{-1}\sigma_j} \right)^2\;,
\end{equation}

\noindent where we use the lensing equation (Eq~\ref{eq:lens}) to
compute the source position $\beta_j$ of the observed image $j$,
$\langle \beta\rangle$ is the barycentre of the $\beta_j$, $\mu_j$ is
the magnification for image $j$, and $\sigma_j$ is the observational
error at the position of image $j$.

Fortunately, we detect no systematics in the recovery.  The error
between the recovered and the input density maps is about 1\% on
average in the hexagon. The most outstanding result is that the
algorithm converged towards the best fit region with only $3\times 10^8$
$\chi^2$ in less than 10 hours on a 2.4~Ghz processor, i.e. it performed more than 9000
$\chi^2/s$ on average and computed only 1.17 $\chi^2$ point per
dimension.  Note that a standard gradient method would have required
at least 3 $\chi^2$ points per dimension to find a minimum i.e.
$3^{122} = 10^{58}$ $\chi^2$ points.

\subsection{Multi-scale model and priors}

\begin{table*}
	\centering
	\caption{\label{tab:scaleprior} \textsc{Priors the multi-scale model}}
	\begin{tabular}{cccccccc}
		\hline
		\hline
		Nb. clumps & $\rcrc$ & Level of splitting & $\sigma_i$
		& $r_{core}^\star$ & $r_{cut}^\star$ & $\sigma_0^\star$
		& $m_K^\star$ \\
		\hline
		120+60 & 3 & 3 & [0,1922] km/s & 0.03'' & [1,40]'' & [10,400] km/s & 16 \\
		\hline
	\end{tabular}
\end{table*}

We consider the multi-scale model of section~\ref{sec:multiscale} to
which we add galaxy-scale clumps. The model then contains a
multi-scale grid of 120 RBF, its smallest element of resolution is
$R_3=12.3''$ and it has a ratio $r_{cut}/r_{core} =3$. We fix the
profile and positions of the RBF. Their weights $\sigma_i^2$ are the
only free parameters. We allow them to vary along a flat prior within the
range 0.1 and 1.9 times the values found in
section~\ref{sec:multiscale} by iterative inversion of
Eq.~\ref{eq:model}. As indicative values, the minimum and
maximum limits we assign are 0 km/s and 1922 km/s. This way, we
prevent the model from wandering too far from the input mass map, but
still give it a substantial freedom.  We also add the 60 galaxy-scale
clumps retained in section~\ref{sec:multiscale}. We assume that {\it
(i)} their position, ellipticity and orientation correspond to the
brightness profile of their associated galaxy, {\it (ii)} their mass
is proportional to the total luminosity of the galaxy, i.e.  they all
have the same M/L ratio and {\it (iii)} they are described by PIEMD
potentials and their PIEMD parameters are linked by the following
scaling relations

\begin{eqnarray}
	\label{eq:scalingrel}
	r_{core} = r_{core}^\star \left( \frac{L}{L^\star}
	\right)^{1/2} \;,
	r_{cut} = r_{cut}^\star \left( \frac{L}{L^\star} \right)^{1/2}
	\;,
	\sigma_0 = \sigma_0^\star \left( \frac{L}{L^\star}
	\right)^{1/4}\;,
\end{eqnarray}

\noindent where $L^\star$, $r_{core}^\star$, $r_{cut}^\star$ and
$\sigma_0^\star$ are respectively the luminosity and the three PIEMD
parameters of a typical early-type galaxy at the cluster redshift.
We consider a cluster galaxy with luminosity $L^\star$
corresponding to an F775W magnitude of 17.54.  We know from previous
studies in this cluster that $r_{cut}^\star \simeq 13''$
\citep{limousin2007b,halkola2007}. Still, we assume a flat prior
between 1'' and 30'' in order to investigate the possible interactions
between the grid and galaxy-scale clumps. Accordingly, we assume
$\sigma_0^\star$ vary along a wide flat prior defined on the range 10
and 400 km/s. Finally, recent studies have shown that the density
profile of early-type galaxies in the field is singular up to
observational limits \citep{koopmans2006,czoske2008}. Therefore, we
fix $r_{core}^\star$ to 0.03'' (i.e. a small value but different from
zero for numerical reasons). In total, this multi-scale model sums 122
free parameters, whose priors are reported in
Table~\ref{tab:scaleprior}.

Finally, note that all the parameters used to build the multi-scale
grid (e.g the input mass map, the hexagonal shape, the hexagon size,
the splitting algorithm, the level of splitting, the
$r_{cut}/r_{core}$ ratio and the surface density threshold) can
potentially affect the inferred strong lensing mass distribution, and
as such must be considered as priors as well.  We further discuss this
point in section~\ref{sec:discussion}. In particular, we look how the
RMS of predicted multiple images improves with the number of splitting
levels.

\subsection{Reference model}

As a reference model, we use a modified version of the Limousin et~al.
model with 2 cluster-scale clumps (for the main and the North-East
halos), 3 galaxy-scale clumps described by individual PIEMD potentials
to model the brightest cluster galaxy (BCG) and Galaxy 1 and Galaxy 2
that strongly affect systems 6, 24 and systems 1, 2 respectively, and
the 60 galaxy-scale clumps of the multi-scale model described by 
scaling relations. In total, the reference model sums 33 free
parameters, whose priors are reported in Table~\ref{tab:ref}.

\begin{table*}
	\centering
	\caption{\label{tab:ref} \textsc{Priors on the reference
	model}}
	\begin{tabular}{cccccccc}
		\hline
		\hline
		ID & R.A. & Decl. & $e$ & $\theta$ &
		$r_{core}$(kpc) &
		$r_{cut}$(kpc) &
		$\sigma_0$(km/s) \\
		\hline
		Clump 1 & [-15,15]& [-15,15] & [0.1,0.55] &
		[0,180] & [30,150] & 1500 & [1000,1700] \\
		Clump 2 & [-90,-35] & [5,79] & [0.4,0.9] & 
		[0,180] & [25,90] & 500 & [300,650] \\
		BCG & [-10,10] & [-10,10] & [0,0.6] & 
		[0,180] & [0.1,10] & [9,550] & [300,680] \\
		Galaxy 1 & 49.0 & 31.5 & [0,0.9] & [0,180] & 
		[0.1,30] & [9,180] & [150,280] \\
		Galaxy 2 & [-49,-45] & [27,35] & [0,0.9] &
		[0,180] & [0.1,20] & [4,190] & [200,580] \\
		$L^\star$ elliptical galaxy & \ldots & \ldots & 
		\ldots & \ldots & 0.15 & [20,60] & [150,280] \\
		\hline
		\multicolumn{8}{l}{\textsc{Notes~:} Coordinates are given in arcseconds with
		respect to the BCG (R.A. = 13:11:29 Decl. = -01:20:27). } \\
		\multicolumn{8}{l}{The ellipticity $e$ is the one of the mass
	distribution, expressed as $a^2 - b^2 / a^2 + b^2$.}
	\end{tabular}
\end{table*}

\subsection{Strong lensing constraints in Abell 1689}

\begin{table*}
	\center
	\caption{ \label{tab:mult}
	\textsc{Multiply imaged systems used as constraints}
	}
	\begin{tabular}{ccccccc}
		\hline
		\hline
		ID & R.A. & Decl. & $\langle\chi^2\rangle$ & $z_{spec}$ &
		\parbox{3.7cm}{\center Image plane rms (arcsec)\\
		Multi-scale model} & 
		\parbox{3.7cm}{\center Image plane rms (arcsec)\\
		Reference model} \\
		\hline
{\bf  1...............}& & &       \bf{1.91} & \bf{3.0} & \bf{0.71} & {\bf 0.67}  \\
      1.1............. & 13:11:26.44 & -1:19:56.37 & 0.96 & &  0.99 & 0.94  \\
      1.3............. & 13:11:29.76 & -1:21:07.31 & 0.96 & &  0.09 & 0.16 \\
{\bf  2...............}& & &      \bf{1.28} & \bf{2.5} & {\bf 0.28} & {\bf 0.59}  \\
      2.1............. & 13:11:26.52 & -1:19:55.07 & 0.64 & &  0.40 & 0.78 \\
      2.4............. & 13:11:29.80 & -1:21:05.95 & 0.64 & &  0.08 & 0.29 \\
{\bf  4.1.............}& & &      \bf{5.07}&{\bf 1.1} & {\bf 0.22} & {\bf 0.67}  \\
      4.11............ & 13:11:32.16 & -1:20:57.33 & 1.23 & &  0.27 & 0.85 \\
      4.12............ & 13:11:30.51 & -1:21:11.90 & 0.90 & &  0.25 & 0.89 \\
      4.13............ & 13:11:30.75 & -1:20:08.01 & 1.64 & &  0.14 & 0.19 \\
      4.14............ & 13:11:26.28 & -1:20:35.06 & 1.30 & &  0.21 & 0.50 \\
{\bf  4.2.............}& & &     \bf{1.33}& {\bf 1.1} & {\bf 0.18} & {\bf 1.08}  \\
      4.21............ & 13:11:32.10 & -1:20:58.31 & 0.66 & &  0.17 & 1.29 \\
      4.22............ & 13:11:30.65 & -1:21:11.41 & 0.66 & &  0.18 & 0.83 \\
{\bf  5...............}& & &     \bf{4.35}& {\bf 2.6}  & {\bf 0.36} & {\bf 1.01}  \\
      5.1............. & 13:11:29.06 & -1:20:48.41 & 2.18 & &  0.49 & 2.18 \\
      5.3............. & 13:11:34.11 & -1:20:20.87 & 2.18 & &  0.15 & 0.49 \\
{\bf  6...............}& & &      \bf{1.73}& \bf{1.1}  & {\bf 0.17} & {\bf 0.31}  \\
      6.1............. & 13:11:30.75 & -1:19:37.90 & 0.87 & &  0.21 & 0.40 \\
      6.2............. & 13:11:33.34 & -1:20:11.97 & 0.87 & &  0.11 & 0.16 \\
{\bf  7...............}& & &      \bf{1.93}&{\bf 4.8 }  & {\bf 0.16} & {\bf 0.23}  \\
      7.1............. & 13:11:25.44 & -1:20:51.52 & 0.96 & &  0.19 & 0.23 \\
      7.2............. & 13:11:30.67 & -1:20:13.79 & 0.96 & &  0.12 & 0.23 \\
{\bf  10.1............}& & &     \bf{0.84}& {\bf 1.8 }  & {\bf 0.10} & {\bf 0.27}  \\
      10.11........... & 13:11:33.96 & -1:20:50.99 & 0.42 & &  0.09 & 0.28 \\
      10.12........... & 13:11:28.05 & -1:20:12.28 & 0.42 & &  0.10 & 0.26 \\
{\bf  10.2............}& & &     \bf{0.68}& {\bf 1.8 }  & {\bf 0.08} & {\bf 0.18}  \\
      10.21........... & 13:11:33.95 & -1:20:51.54 & 0.34 & &  0.08 & 0.20 \\
      10.22........... & 13:11:28.08 & -1:20:11.84 & 0.34 & &  0.08 & 0.17 \\
{\bf  10.3............}& & &   \bf{0.77} &  {\bf 1.8 }  & {\bf 0.08} & {\bf 0.19}  \\
      10.31........... & 13:11:33.96 & -1:20:51.53 & 0.39 & &  0.08 & 0.20 \\
      10.32........... & 13:11:28.08 & -1:20:12.24 & 0.39 & &  0.09 & 0.18 \\
{\bf  24.1............}& & &    \bf{3.34}&  {\bf 2.6 }  & {\bf 0.17} & {\bf 0.52}  \\
      24.11........... & 13:11:29.18 & -1:20:56.04 & 1.63 & &  0.18 & 0.79 \\
      24.13........... & 13:11:30.29 & -1:19:33.86 & 0.78 & &  0.20 & 0.25 \\
      24.14........... & 13:11:33.71 & -1:20:19.82 & 0.94 & &  0.13 & 0.33 \\
{\bf  24.2...........}& & &    \bf{2.92}&   {\bf 2.6 }  & {\bf 0.18} & {\bf   0.40}  \\
      24.21........... & 13:11:29.22 & -1:20:55.28 & 0.77 & &  0.15 & 0.38 \\
      24.23........... & 13:11:30.25 & -1:19:33.26 & 1.30 & &  0.25 & 0.55 \\
      24.24........... & 13:11:33.69 & -1:20:18.80 & 0.85 & &  0.12 & 0.21 \\
      \hline
      \end{tabular}
\end{table*}

From \citet{limousin2007b} we select a catalogue of 28 images in 12
systems of multiple images (see Table~\ref{tab:mult}). We will use the
rest of the images later to check for model predictability and
over-fitting. Selected
images must have a measured spectroscopic redshift (Richard et~al.,
prep.), or present at least two star-forming regions clearly visible
in all the counter-images, in order to avoid identification mistakes.
We also remove images whose light is blended with the light of nearby
galaxies. To get accurate measurements of the images positions, we fit
a Gaussian profile to their light distribution with the \textsc{IRAF}
task \textsc{imexam}.  Because our selection mostly contains compact
images, we measure a Gaussian width of 0.13'' on the position
measurements. We use this positional error for all the images in the
lensing $\chi^2$ (Eq.\ref{eq:chi2}). Since we only use the image
position as a constraint, we sum 32 constraints.  

\subsection{Computational considerations}

On a 2.4~Ghz processor, the Bayesian estimation of the 122 free
parameters took 15 days to produce about 20000 MCMC samples. Although
it could be considered as quite computational intensive, considering
the number of free parameters and the fact that we not only find the
best fit region but also explore the parameter space in its
neighbourhood, it is in fact very efficient. In the last section, we
discuss some issues related to the number of samples.

\subsection{Results}

\subsubsection{Images prediction}

Results confirm the ability of multi-scale models at being used as
lens models. Indeed, the RMS per system and per image reported in
Table~\ref{tab:mult} and shown in Fig.~\ref{fig:mult_pred}a highlight
the good precision we obtain. The
RMS averaged over all the systems is 0.28''.  In contrast, the
reference model optimised with the same catalogue of images produces a
mean RMS of 0.54''. For all the systems unless System 1, RMS with the
multi-scale model are lower than RMS with the reference model.
System 1 has a slightly larger RMS because the galaxy-scale clump
located 3.2'' to the West may not fit the same scaling relations as
the other galaxy-scale clumps. Indeed, if we assume that 2 galaxies do
not follow the same scaling relations, there is no perfect solution for
the scaling relations parameters.  The images producing the largest
$\chi^2$ bias the scaling relations in their favor. The fit of 
images with a lower $\chi^2$ but close to galaxies following
other scaling relations worsens.  In this respect, note that for
System 1, $\chi^2 = 1.91$ and RMS$=0.71''$ whereas for System 24.1
$\chi^2=3.34$ and RMS$=0.17''$. Both systems are at the same distance
of a galaxy-scale clump.  System 24.1 is therefore likely biasing the
scaling relations in its favor.

\begin{figure} 
	\centering
	\includegraphics[width=\linewidth]{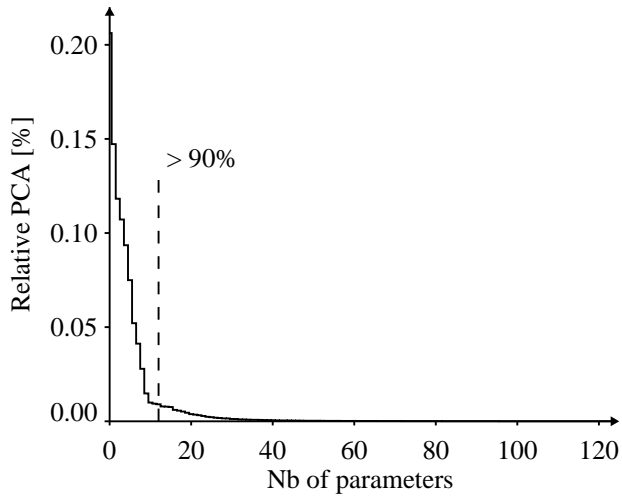} 

	\caption{\label{fig:pca} Relative contributions of the PCA
	eigenvalues to the total variance of the samples distribution
	in the our parameter space with 122 dimensions.  Note that
	90\% of the variance is reproduced with only 12 effective
	parameters.	}

\end{figure}

Given the large number of free parameters compared to the number of
constraints, we could have expected better RMS and smaller $\chi^2$.
Indeed, it is usually accepted that models with more free parameters
than constraints allow an infinity of perfect solutions. Since our
solutions are not perfect, it means that the number of useful
effective parameters in our model is actually
lower than the number of constraints. To estimate this number, we
analyse the distribution of the MCMC samples in the parameter space by
means of the Principal Component Analysis (PCA) technique.
Fig.~\ref{fig:pca} shows that 90\% of the variance of the distribution
is reproduced with only 12 effective parameters out of the 122 `real'
parameters. With the reference model, 90\% of the variance is
reproduced with only 10 effective parameters out of the 33 `real' ones. In
both cases, the number of effective parameters is lower than the 32
constraints, hence the non perfect fit. Though very interesting,
investigating the physical meaning of these effective parameters is
out of the scope of this paper.

\begin{figure} \centering
	\includegraphics[width=\linewidth]{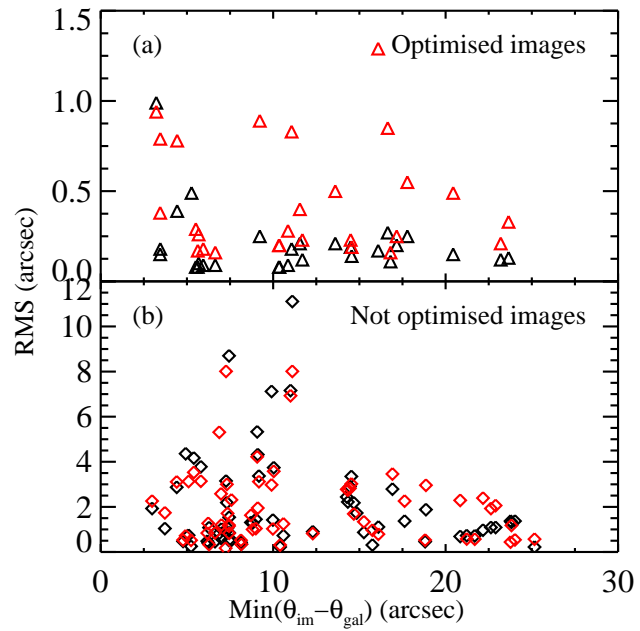}

	\caption{\label{fig:mult_pred} {\bf (a)} Image plane
	RMS of the images used as constraints in function of their
	distance to the closest galaxy. {\bf In black:} RMS obtained
	with the multi-scale model. {\bf In red:} RMS with the
	reference model. {\bf (b)} RMS produced by the images
	kept for the over-fitting check. The two subsets give similar
	RMS, indicating no particular sign of over-fitting with the
	multi-scale model. } 

\end{figure}

As a definite confirmation that our multi-scale model is not
under-constrained and do not over-fit the data, we apply the cross-checking
technique~: with both the multi-scale and the reference models, we
predict the images positions for the part of the images
catalogue not used as constraints, and compare the RMS between
predicted and observed positions. This way, if a model is
biased towards the subset of images used as constraints, it should be
unable to give accurate predictions. Fig.\ref{fig:mult_pred}b shows
the RMS given the two models for the not optimised part of the images
catalogue. On average, we find an RMS
of 3.32'' for the multi-scale model and an RMS of 3.49'' for the reference
model.  Since the two models give similar predictions, we
conclude that the multi-scale model does not over-fit the data.

In addition, Fig.~\ref{fig:mult_pred}b also shows that when images
get closer to galaxies their RMS increases. This increasing RMS
again suggests that galaxy-scale clumps do not perfectly fit the imposed
scaling relations. There must exists a scatter in the scaling
relations that images seem to be sensible to, and that should be
included in our future models.

\subsubsection{Convergence, deviation and shear maps}

\begin{figure*} 
	\centering 
	\includegraphics[width=\textwidth]{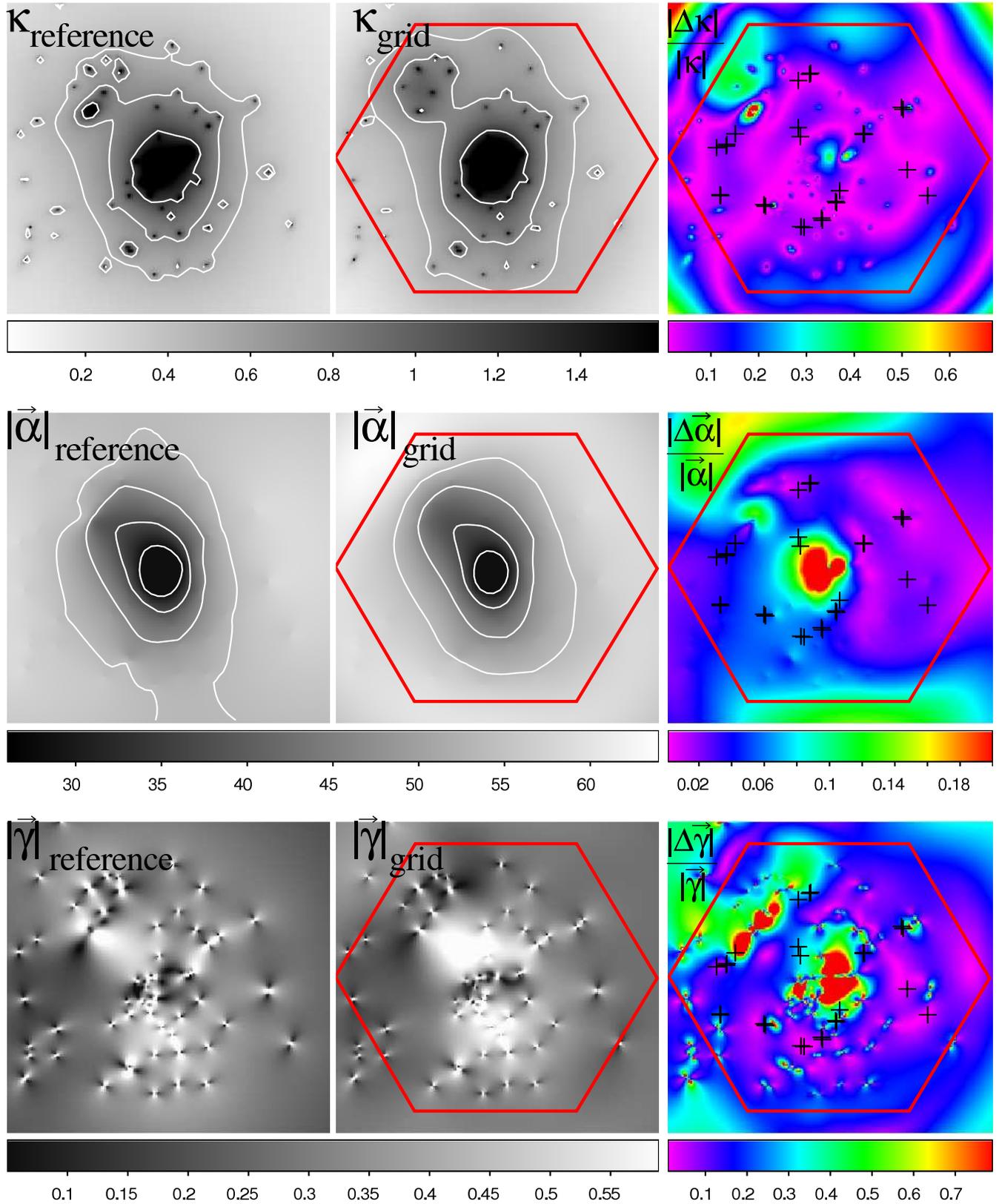}

	\caption{\label{fig:resSL} Comparison between maps obtained
	with the reference model (left column) and the
	multi-scale model (middle column). The color images correspond
	to the relative difference
	$|\mathrm{Reference}-\mathrm{grid}|/|\mathrm{Reference}|$
	between the two maps in grey. All the maps are computed for a
	source plane at redshift $z_S =  3$.  The red hexagon marks the
	limit of the multi-scale grid beyond which the maps become
	meaningless. The black crosses in the color images mark the
	positions of the images used as constraints.} 

\end{figure*}

Fig.~\ref{fig:resSL} compares mean convergence, deviation and
shear maps produced with the reference and the multi-scale models. To
produce these maps, we compute the convergence, deviation and shear
maps for each mass model of the MCMC chain, and compute the mean maps
by averaging the values of each pixel of each map. First of all, note
that the two convergence maps are very similar. The better RMS
obtained with the multi-scale model does not originate from any
particular missing clump in the reference model. The
mass enclosed at the Einstein radius $M(<45'')$ are also very similar
with less than 1\% difference. \emph{In other words, the ``mass
follows light'' assumption in traditional modelling holds}.
Nonetheless, the better RMS obtained with the multi-scale model
attests a significantly higher degree of flexibility. In particular,
we note that in the convergence map, the North-East clump, at
intermediate scale between cluster and galaxy scale, looks
smoother and more detached from the main clump than in the convergence
map produced by the reference model. The deviation maps produced by
the two models are also very similar with less than 5\% of difference
at the images position. Finally the errors in the shear maps are
mostly lower than 10\% at the images position. This is currently below
what we can observationally constrain by measuring the ellipticity of
multiple images. Consequently, the better RMS with the multi-scale
model is merely due to its large number of free parameters, which allow a refined modelling of the mass distribution
irregularities at intermediate scale between cluster and galaxy scale.

\subsubsection{Error mass map}

\begin{figure} 
	\center 
	\includegraphics[width=\linewidth]{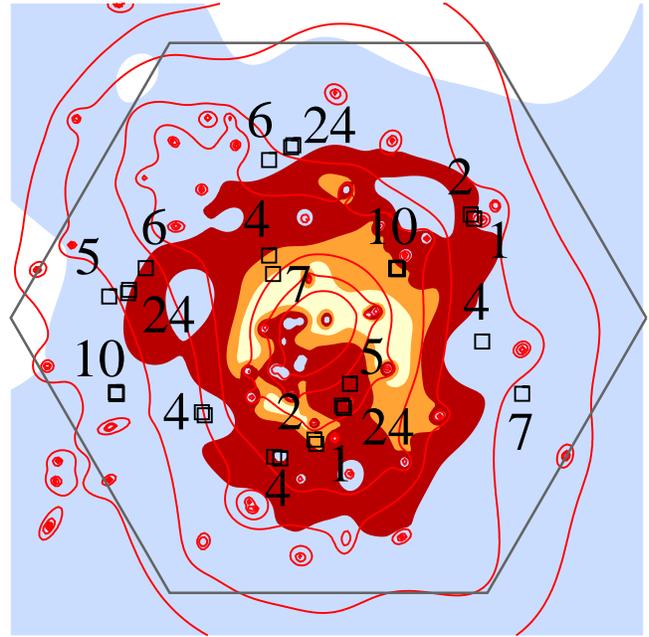}

	\caption{ \label{fig:emass} Map of S/N ratio. Coloured
	contours bound regions with S/N greater than 300, 200, 100 and
	10, the highest S/N region being at the centre.  Red contours
	are mean iso-mass contours from 5 to $0.77\times10^{12}\
	M_\odot/\mathrm{arcsec}^2$ by steps of 1 and
	$0.5\times10^{12}\ M_\odot/\mathrm{arcsec}^2$ for the three
	outer contours. This map has been computed with 20000 MCMC
	samples. Black boxes mark the positions of the multiple images
	used to constrain the mass distribution. }

\end{figure}

In addition to the mean convergence map, the Bayesian approach also
allows to compute the standard deviation and the $S/N$ convergence
maps.  Fig.~\ref{fig:emass} shows that the mass distribution is
estimated with high confidence, since the $S/N$ is everywhere larger
than 10 inside the hexagon, i.e. less than 10\% error.  This means that
the correlations between the RBF parameters, highlighted by the PCA
technique, must strongly restrict the range allowed by the priors,
hence the small error. Furthermore, the smooth iso-mass contours
confirm the ability of multi-scale models at producing smooth mass
maps.

\section{Discussion}
\label{sec:discussion}

In the previous section, we have worked at demonstrating the strength
of our model. We are now going to highlight some critical aspects
which will give the reader a more complete view of the multi-scale
models, but also of ``non-parametric'' models in general, for which the effects
of grid parameters on the final results are rarely investigated in
details. In this section, we particularly investigate two parameters~:
{\it (i)} the $\rcrc$ ratio defining the RBF concentration and {\it
(ii)} the level of splitting. We do not investigate the threshold
parameter because we only aim at oversampling  strong lensing regions.
Therefore hereafter, in all the models, the surface density threshold
is set equal to $\Sigma_{crit} =
2\times10^{10}\;M_\odot/\mathrm{arcsec}^2$. 

To begin with, we build 5 multi-scale models similar to the one used
so far, but with different $\rcrc$ ratios and
levels of splitting. Some models built with 3 levels of
splitting have 127 clumps because the grid has been
shifted by 6'' with respect to the input mass map. We found that this
shift does not affect the final results. We optimise each model with
our catalogue of 28 images and report the mean RMS and mean likelihood
$\langle \log(L) \rangle = -0.5\;\langle \chi^2 \rangle$ in
Table~\ref{tab:models}.  Note also that less MCMC samples have been
gathered with models A and C in order to evaluate how the number of
MCMC samples affects the error estimation.

\begin{table*} 
	\center 
	\caption{\textsc{List of multi-scale models used in this section}} 
	\begin{tabular}{cccccccc} \hline \hline
		ID & $r_{cut}/r_{core}$ ratio  & Nb of splitting
		levels & Nb of clumps & Nb samples & $M_{gal}/M_{tot}$
		ratio & Mean Image plane
		rms & $\langle \log(L) \rangle$ \\
		\hline 
		A & 2   &   3    & 127  & 439 & 14\% & 0.23''  & -14.8 \\
		B & 3   &   3    & 120  & 20920 & 13\% & 0.27''  & -13.4 \\ 
		C & 4   &   3    & 127  & 369  & 12\% & 0.33'' & -28.2 \\
		D & 10  &   3    & 127  & 25029 & 11\% & 0.86'' & -228.2 \\ 
		E & 3   &   4    & 318  & 38399 & 5\% & 0.22'' & -13.1 \\ 
		Reference & - & - & - & 1000  & 22\% & 0.57'' & -59.4 \\
		\hline 
	\end{tabular} 
	\label{tab:models} 

\end{table*}

\subsection{The $\rcrc$ ratio}

\begin{figure}
	\centering
	\includegraphics[width=\linewidth]{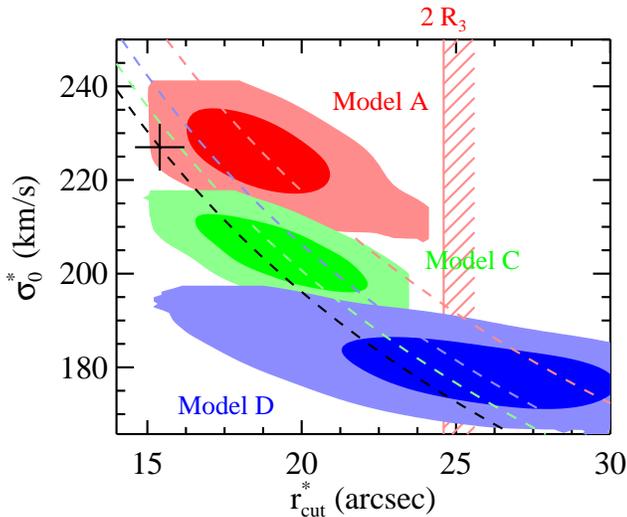}

	\caption{\label{fig:cont_rcut}
	Confidence intervals at 68\% and 99\% of the scaling relations
	parameters $r_{cut}^\star$ and $\sigma^\star$ obtained with
	models A, C and D. Dashed lines show the curves of constant
	M/L ratio within a 60'' aperture (i.e. the mean distance from
	multiple images to galaxy-scale clumps). In black is shown the
	same curve but for the Limousin et~al.  model. The
	vertical dashed limit at $2R_3 = 24.6''$ marks the grid
	resolution of model A. The grid resolution of models C and D
	at $4R_3$ and $10R_3$ respectively, are beyond the abscissa
	scale limit of this plot. This figure shows that galaxy-scale
	clumps efficiently increase the grid resolution. }

\end{figure}

\begin{figure}
	\centering
		\includegraphics[width=\linewidth]{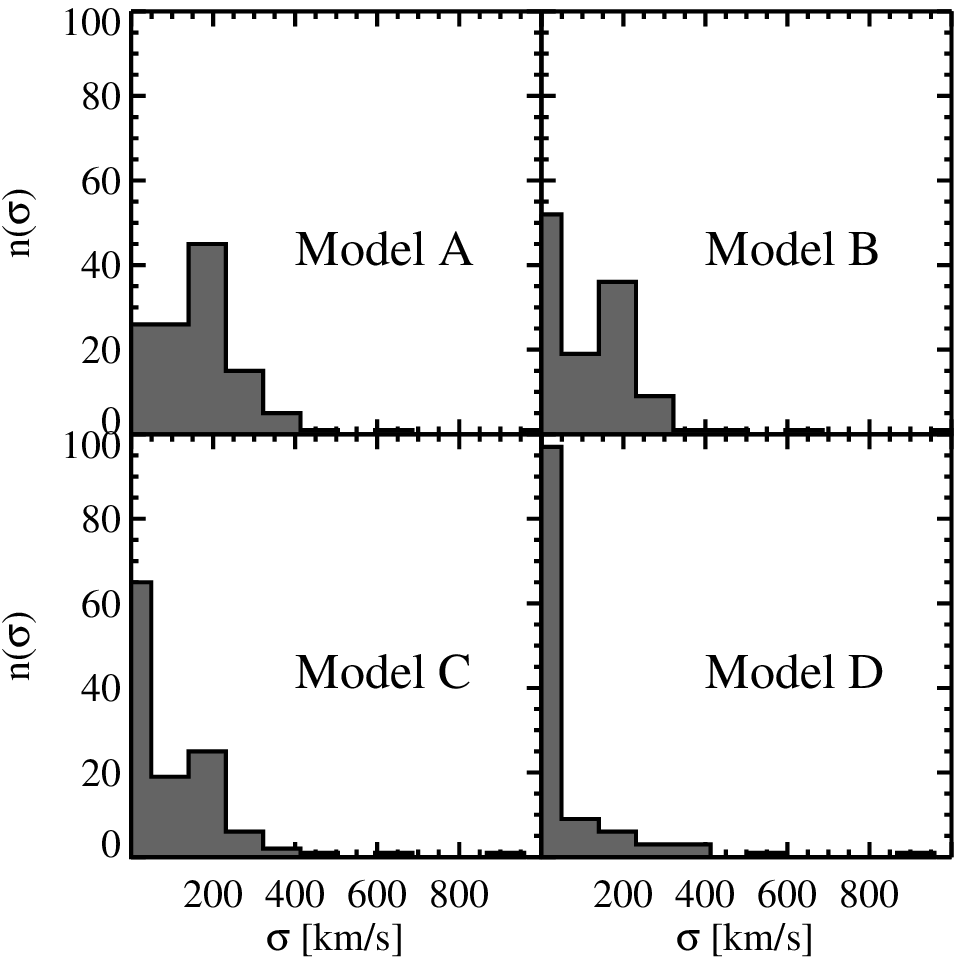}

	\caption{\label{fig:n_sigma}
	Histograms of the RBF velocity dispersions in models A, B, C
	and D, as distributed before strong lensing optimisation (these
	distributions slightly shrink after optimisation, i.e.  small
	values get larger and large values get smaller). Note
	that multi-scale models are mainly made of RBF with small
	velocity dispersion. As RBF get more extended, more
	RBF are assigned small velocity dispersions. }

\end{figure}

The $\rcrc$ ratio characterises the concentration of the RBF. Models
with small ratios are more flexible because their RBF are more
independent.  Table~\ref{tab:models} shows that such models better fit
multiple images but can also over-fit them. In contrast, models with
larger ratios produce more extended and overlapping RBF, which worsens
the fit to multiple images. For instance, model A with $\rcrc=2$ (i.e.
with very concentrated RBF) gives a better fit than model D, with
$\rcrc = 10$.  If we calculate their likelihood ratio we find
$\log(L_D/L_A) = 113$, which means that model D is clearly ruled out. 

In addition to the fit quality, the $\rcrc$ ratio affects the scaling
relations parameters $r_{cut}^\star$ and $\sigma_0^\star$.
Fig.~\ref{fig:cont_rcut} shows that models with large $\rcrc$ ratio
(i.e. extended RBF) induce large $r_{cut}^\star$ and small
$\sigma_0^\star$. To understand this behaviour, Fig.~\ref{fig:n_sigma}
shows that in such models, most of the RBF have a small velocity
dispersion, i.e.  little weight. The mass distribution must then be
very smooth with only a few very extended and effective RBF. The
$r_{cut}^\star$ extension parameter becomes large, in order for the
galaxy-scale clumps to accommodate the multiple images and compensate
for the lack of effective RBF in their neighbourhood. Besides, whatever
the $\rcrc$ ratio, Fig.~\ref{fig:cont_rcut} shows that the curves of
constant M/L ratio agree at 2$\sigma$, and Table~\ref{tab:models}
shows that for a given level of splitting, the $M_{gal}/M_{tot}$
ratios are similar. In conclusion, the $\rcrc$ ratio affects the
extension of the galaxy-scale clumps but do not affect their mass. 

\begin{figure}
	\centering
	\includegraphics[width=\linewidth]{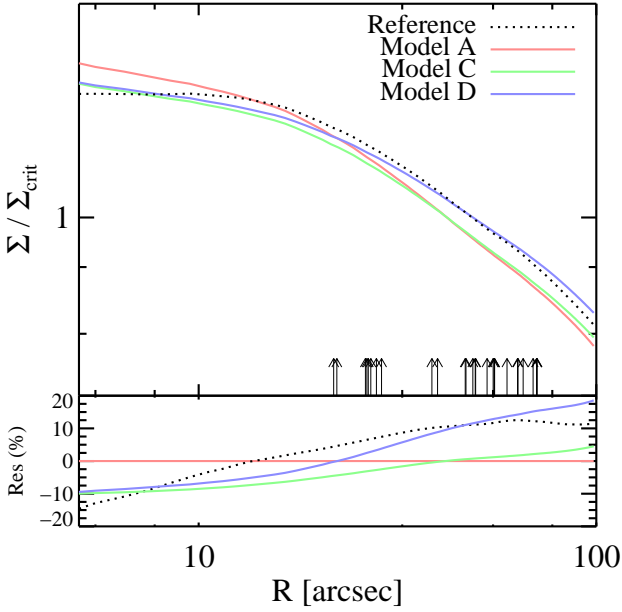}

	\caption{\label{fig:massprof_ratio}
	Convergence profiles obtained with models A, C and D. Larger
	$\rcrc$ ratios make the profile shallower.  The vertical arrows
	radially locate the strong lensing constraints.  } 

\end{figure}

As already noticed by \citet{diego2007}, extended RBF produce a mass
sheet excess at large radius.  Fig.~\ref{fig:massprof_ratio} shows
that model D with a large $\rcrc$ ratio produces a density
profile with a shallower slope than model A. Since with larger ratios,
the RBF become more extended, their cut-off radius is pushed out to
larger radius. Their isothermal slope is maintained over a wider range
in radius, making the total density profile shallower. 

\subsection{The level of splitting}

\begin{figure}
	\centering
	\includegraphics[width=\linewidth]{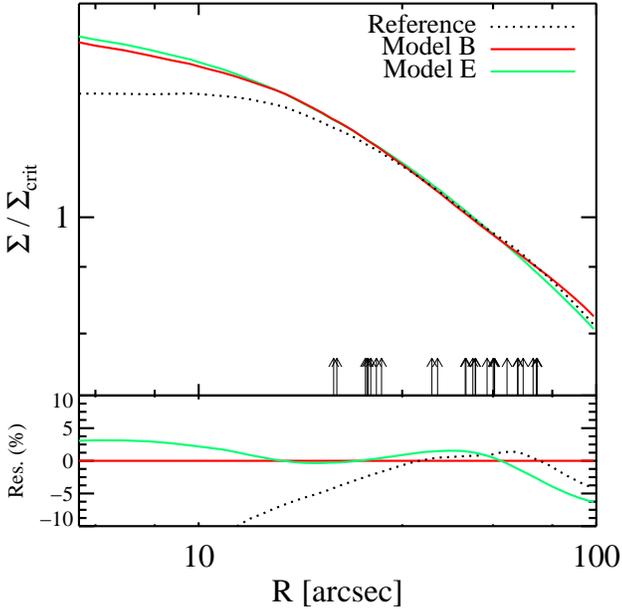}

	\caption{\label{fig:massprof_levels_wgals}
	Density profile comparison between multi-scale models obtained
	with 3 and 4 levels of splitting, hence made of 120 and 318
	RBF respectively, and 60 galaxy-scale clumps. The vertical
	arrows radially locate the strong lensing constraints. In this
	region, the density profiles are similar. Outside, their are
	mostly driven by the priors. } 

\end{figure}

\begin{figure}
	\centering
	\includegraphics[width=\linewidth]{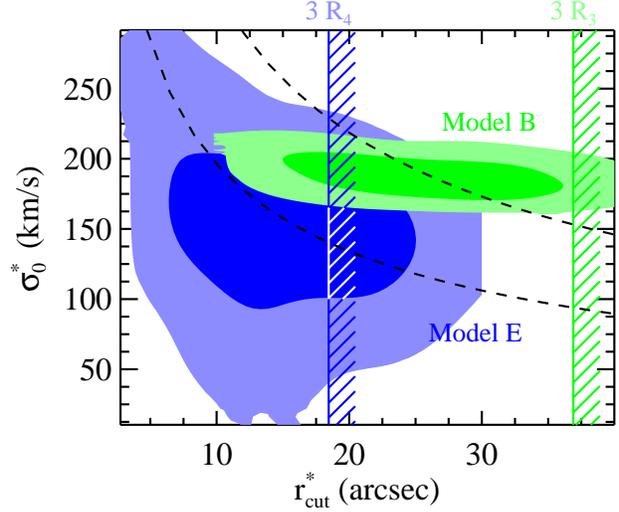}

	\caption{\label{fig:cont_rcut_levels}
	Confidence levels at 68\% and 99\% of the scaling parameters
	$r_{cut}^\star$ and $\sigma_0^\star$ recovered with
	multi-scale models B (green) and D (blue). Dashed lines show
	the curves of constant M/L ratio within a 60'' aperture.
	Dashed limits at $3R_3$ and $3R_4$ mark the extension of the
	smallest RBF for models B and D respectively. Note
	that with 4 levels, the galaxy-scale clumps are completely
	unconstrained because they directly compete with the RBF. }

\end{figure}

Another prior involved in the building of the grid is the level of
splitting. Fig.~\ref{fig:massprof_levels_wgals} shows that in contrast
to the $\rcrc$ ratio, the level of splitting does not affect the slope
of the density profile. The profiles obtained with models B, E and the
reference model are very similar. In addition, Table~\ref{tab:models}
shows that models B and E have similar RMS and likelihood. Therefore,
more levels of splitting is not expressly justified given the data.
Note also that the reference model produces a similar density profile
but has an RMS twice the RMS obtained with models B or E. Similar
density profiles does not necessarily mean similar fit to the data.

Fig.~\ref{fig:cont_rcut_levels} shows that the level of splitting
affects the scaling relations parameter $r_{cut}^\star$. With model E,
$r_{cut}^\star \simeq 15''$ whereas with model B, $r_{cut}^\star
\simeq 25''$. In addition, it shows that with model E, $r_{cut}^\star$
is close to the grid resolution $3R_4=18.4''$, whereas with model B, it is
significantly smaller, since $3R_3 = 36.9''$. It seems therefore that
by increasing the level of splitting, we just replace the galaxy-scale
clumps by the RBF of the grid. The large errors in model E indeed indicate
that the galaxy-scale clumps are not constrained anymore by the data.
Besides, Fig.~\ref{fig:cont_rcut_levels} and Table~\ref{tab:models}
show that the M/L ratio decreases as the level of splitting increases.
The contribution of the galaxy-scale clumps to the total mass
decreases from 13\% in model B to 5\% in model D.  In other words,
with 4 levels of splitting, galaxy-scale clumps do not help anymore in
increasing the model resolution.  The scaling relations inflexibility
might imped them from decreasing in size, to make the model gain in
resolution and improve the fit. Again, one solution could be to introduce a
scatter in the scaling relations. We will explore this solution in a
forthcoming paper.

\subsection{Invariant quantity}

\begin{figure}
	\centering
	\includegraphics[width=\linewidth]{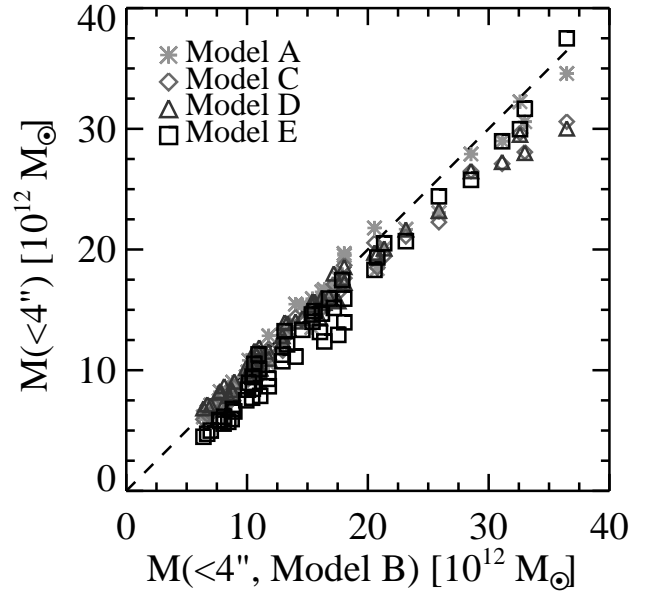}

	\caption{\label{fig:mass_raper}
	Galaxy-scale clumps mass measurements within a 4'' aperture,
	performed on pixelated mass maps resulting from the fit of
	models A, B, C, D and E to the same subset of multiple images.
	Measurements are very consistent with a scatter of about 10\%,
	in agreement with the errors observed on the radial density
	profiles.
	}
	
\end{figure}

So far, we have found that both the $\rcrc$ ratio and the level of
splitting affect the estimation of the scaling relations parameters
$r_{cut}^\star$, $\sigma_0^\star$, as well as the M/L ratio of the
galaxy-scale clumps. In this context, studying the galaxy-scale clumps
physical properties seems a little risky. Nevertheless, it is possible
to get more reliable values by making measurements directly on the
surface density maps. In Fig.~\ref{fig:mass_raper}, we compute the
aperture mass of galaxy-scale clumps with \textsc{Sextractor}
\citep{bertin1996} and pixelated mass maps obtained for models A,
B, C, D and E. We choose \textsc{Sextrator} because it affords a
multi-threshold algorithm to assign the mass in a pixel to the most
credible of two nearby galaxies.  We find that the masses enclosed in
a 4'' aperture (i.e.  the smallest distance between two nearby
galaxy-scale clumps) are almost unaffected by the grid parameters. The
scatter is of the order of 10\%, in agreement with the errors found in
the density profiles.  Estimating the galaxy-scale clumps properties
directly from the surface density maps seems therefore a more reliable
and better constrained solution (given the priors) than simply
considering the $r_{cut}^\star$ and $\sigma_0^\star$ modelling
parameters.

\subsection{Errors analysis}

\begin{figure}
	\centering
	\includegraphics[width=\linewidth]{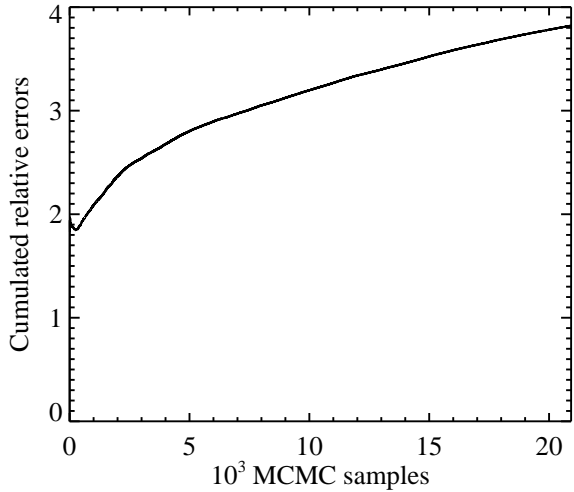}

	\caption{\label{fig:err_total}
	Evolution of the cumulated relative error of the parameters of
	model B in function of the number of MCMC samples. The error
	does not stabilise even after more than 20000 samples. }

\end{figure}

\begin{figure}
	\centering
	\includegraphics[width=\linewidth]{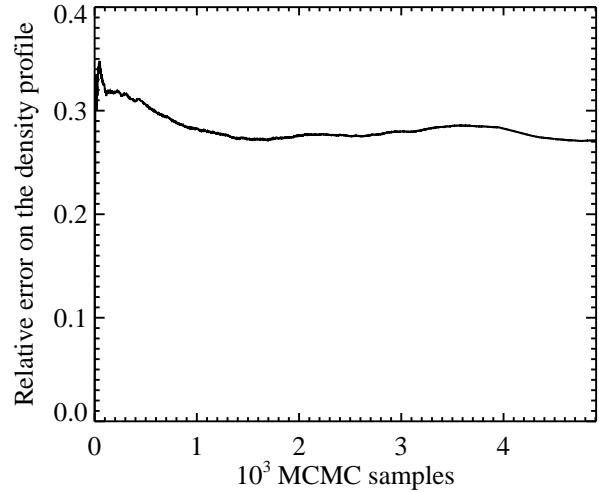}

	\caption{\label{fig:err_massprof}
	Evolution of the relative error on the radial density profile of
	model B in function of the number of MCMC samples. The error
	stabilises soon after 1500 samples. }

\end{figure}

As already stressed, given the size of the parameter spaces, it is
encouraging to see our Bayesian MCMC sampler converging to the best
fit region in less than 5000 samples. However,
Fig.~\ref{fig:cont_rcut_levels} shows that the error bars increase
between model B and D. Since model D sums twice as many samples as
model B, it seems that the estimation of the errors depends on the
number of computed samples. To study the evolution of the errors
estimation with the number of MCMC sample with the number of MCMC
sampless, we
define the cumulated relative error quantity computed over all the
free parameters of the model as 

\begin{equation}
	\mathrm{Cumulated\ parameters\ error} = \sqrt{ \sum_i \left( \frac{\sigma[X_i]}{E[X_i]} \right)^2 }
\end{equation}

\noindent where $\sigma[X_i]$ and $E[X_i]$ are the standard deviation
and the mean value of the $X_i$ MCMC random variable for parameter $i$.
Fig.~\ref{fig:err_total} shows that this error has not converged even
after more than 20000 samples. Similarly, we define the error on the
density profile as

\begin{equation}
	\mathrm{Density\ profile\ error} = \sqrt{ \int_0^{100} \left( \frac{\sigma[S_i]}{E[S_i]} \right)^2
	\mathrm{d}R }
\end{equation}

\noindent where $S_i$ is a random variable for the radial density profile
integrated on the range 0 to 100''. In contrast to the error on the
free parameters, Fig.~\ref{fig:err_massprof} shows that the error on
the density profile converges after only 1500 samples.  Therefore,
accurate errors on the mass distribution are fast to compute, whereas
errors on the parameters of the model, such as $r_{cut}^\star$ or
$\sigma_0^\star$ for instance, are difficult to estimate accurately
and might actually be underestimated.  

\section{Conclusion}

In this paper, we present a multi-scale model sufficiently flexible to
reproduce the observed systems of multiple images with high accuracy,
but also robust against over-fitting. The model combines a grid of
radial basis functions (RBF), and galaxy-scale clumps hosting cluster
member galaxies, described by PIEMD potentials and scaling relations. 

We apply this model to the galaxy cluster Abell~1689. We constrain the
model with a subset of multiple images extracted from the
\citet{limousin2007b} catalogue. We obtain an RMS between observed and
predicted positions of 0.28'' in the image plane, i.e.  about half the
RMS obtained with a slightly modified version of the Limousin et~al.
model (the reference model). We confirm predictability of our multi-scale
model by cross-checking the optimised model with the part of the
images catalogue not used as constraints. We find similar RMS with the multi-scale and the
reference models. This confirms that we do not overfit the data,
despite the large number of free parameters. We propose a Principal
Component Analysis (PCA) technique to estimate the effective number of
free parameters. We find that our multi-scale model with 122
parameters contains only 12 effective parameters. Our multi-scale
model produces smooth mass maps and radial density profiles.

Then, we compare the convergence, deviation and shear maps obtained
both with the multi-scale and the reference models. We find no major
difference between the two sets of maps.  This means that the better
RMS is due to minor changes allowed by the flexibility of the
multi-scale model at intermediate scale between cluster and galaxy
scale. We also note that the galaxy-scale clumps are better integrated
to the North-East cluster-scale clump with the multi-scale grid than
with the traditional reference model. The better modelling of this
cluster-scale component, embedded in projection in the main cluster
core, thus illustrates how multi-scale models can reproduce irregular
mass distributions. 

Finally, we study how changes on the grid parameters affect the
density profile, the scaling relations parameters $\sigma_0^\star$,
$r_{cut}^\star$, and the M/L ratio of the galaxy-scale clumps. We find
that galaxy-scale clumps efficiently increase the resolution of the
grid of RBF without modifying the density profile. Excessively raising
the level of splitting in order to increase the grid resolution is
thus not expressly justified. We also find that the scaling relations
parameter $r_{cut}^\star$, related to the galaxy-scale clumps
extension, decreases as the RBF get more
concentrated. This degeneracy leads to several models producing the
same radial density profile.  Fortunately, the likelihood is also
significantly affected, hence allowing an effective model selection.
In spite of this degeneracy between the model's parameters, we propose
a reliable solution to measure the galaxy-scale clump mass into an
aperture directly on the mass maps produced by the optimised models.
Indeed, we note that the model's parameters degenerate, but the
resulting mass distributions are little affected by the priors on the
grid. 

This work raises the issue of accurately measuring substructure
properties in galaxy clusters
\citep{halkola2007,natarajan2007a,smith2008}. The degeneracy between
galaxy-scale and cluster-scale mass components can potentially lead to
spurious conclusions. In traditional modelling of observationally
quiet clusters, whatever the priors on the adopted profile for the
cluster-scale component (e.g. PIEMD or NFW profile), the substructure
properties are little affected, given the constraints.  However, in
unrelaxed perturbed clusters, pursuing this study with the same
technique appears to be more risky. Our multi-scale model as described
above offers an interesting solution. 

In terms of computation time, this method is still slow.  We could
make it faster by restricting the number of gradients and Laplacians
calculated per image to solely the most significant.  However, this
has to be treated with care since the sum of negligible gradients can
become significant. Missing such gradients could lead to spurious
results.  Look for instance \citet{deb2008} to understand how complex
it is to accurately select the relevant clumps for gradient
calculation.

Multi-scale models open interesting avenues for the modelling of
galaxy clusters. Indeed, today standard parametric methods are facing
their own limitations. New lensing results reveal irregular or
merging clusters \citep[e.g.][]{bradac2006,jee2007,mahdavi2007}.
Multi-scale models with RBF afford the robustness of parametric
methods as well as the flexibility of non-parametric methods.
Since we use analytical potentials, the extension of the method to
weak lensing is straightforward. We are currently working at combining
strong and weak lensing signal to extend the accuracy of our modelling
to larger radius. In inner regions, the strong lensing constraints
will need a dense network of RBF to be reproduced accurately
whereas in the outskirts, a coarser sampling will perfectly fit with
the lower density of weak lensing constraints (see Jullo et al. {\it
in prep}).  This method can also be extended to multiplane lensing,
allowing thus tomographic analysis.

\vspace{1cm}
The authors thank Phil Marshall, Harald Ebeling, Marceau Limousin and
Johan Richard, as well as the referee for enlightening and fruitful
comments. JPK and EJ thank support from CNRS and SL2S grant
06-BLAN-0067 from the Agence National de la Recherche.

\bibliographystyle{/scisoft/share/texmf/mn/mn2e} 
\bibliography{jullo_jpk}

\begin{thebibliography}{}

\bibitem[\protect\citeauthoryear{{Bertin} \& {Arnouts}}{{Bertin} \&
  {Arnouts}}{1996}]{bertin1996}
{Bertin} E.,  {Arnouts} S.,  1996, \aaps, 117, 393

\bibitem[\protect\citeauthoryear{{Blandford} \& {Narayan}}{{Blandford} \&
  {Narayan}}{1986}]{blandford1986}
{Blandford} R.,  {Narayan} R.,  1986, \apj, 310, 568

\bibitem[\protect\citeauthoryear{{Brada{\v c}}, {Clowe}, {Gonzalez},
  {Marshall}, {Forman}, {Jones}, {Markevitch}, {Randall}, {Schrabback} \&
  {Zaritsky}}{{Brada{\v c}} et~al.}{2006}]{bradac2006}
{Brada{\v c}} M.,  {Clowe} D.,  {Gonzalez} A.~H.,  {Marshall} P.,  {Forman} W.,
   {Jones} C.,  {Markevitch} M.,  {Randall} S.,  {Schrabback} T.,    {Zaritsky}
  D.,  2006, \apj, 652, 937

\bibitem[\protect\citeauthoryear{{Brada{\v c}}, {Schneider}, {Lombardi} \&
  {Erben}}{{Brada{\v c}} et~al.}{2005}]{bradac2005}
{Brada{\v c}} M.,  {Schneider} P.,  {Lombardi} M.,    {Erben} T.,  2005, \aap,
  437, 39

\bibitem[\protect\citeauthoryear{{Broadhurst}, {Ben{\'{\i}}tez}, {Coe} \&
  {et~al}}{{Broadhurst} et~al.}{2005}]{broadhurst2005}
{Broadhurst} T.,  {Ben{\'{\i}}tez} N.,  {Coe} D.,    {et~al} 2005, \apj, 621,
  53

\bibitem[\protect\citeauthoryear{{Clowe}, {Brada{\v c}}, {Gonzalez},
  {Markevitch}, {Randall}, {Jones} \& {Zaritsky}}{{Clowe}
  et~al.}{2006}]{clowe2006}
{Clowe} D.,  {Brada{\v c}} M.,  {Gonzalez} A.~H.,  {Markevitch} M.,  {Randall}
  S.~W.,  {Jones} C.,    {Zaritsky} D.,  2006, \apjl, 648, L109

\bibitem[\protect\citeauthoryear{{Coe}, {Fuselier}, {Ben{\'{\i}}tez},
  {Broadhurst}, {Frye} \& {Ford}}{{Coe} et~al.}{2008}]{coe2008}
{Coe} D.,  {Fuselier} E.,  {Ben{\'{\i}}tez} N.,  {Broadhurst} T.,  {Frye} B.,
   {Ford} H.,  2008, \apj, 681, 814

\bibitem[\protect\citeauthoryear{{Czoske}, {Barnab{\`e}}, {Koopmans}, {Treu} \&
  {Bolton}}{{Czoske} et~al.}{2008}]{czoske2008}
{Czoske} O.,  {Barnab{\`e}} M.,  {Koopmans} L.~V.~E.,  {Treu} T.,    {Bolton}
  A.~S.,  2008, \mnras, 384, 987

\bibitem[\protect\citeauthoryear{{Czoske}, {Moore}, {Kneib} \&
  {Soucail}}{{Czoske} et~al.}{2002}]{czoske2002}
{Czoske} O.,  {Moore} B.,  {Kneib} J.-P.,    {Soucail} G.,  2002, \aap, 386, 31

\bibitem[\protect\citeauthoryear{{de Blok}, {Bosma} \& {McGaugh}}{{de Blok}
  et~al.}{2003}]{deblok2003}
{de Blok} W.~J.~G.,  {Bosma} A.,    {McGaugh} S.,  2003, \mnras, 340, 657

\bibitem[\protect\citeauthoryear{Deb, Goldberg \& Ramdass}{Deb
  et~al.}{2008}]{deb2008}
Deb S.,  Goldberg D.~M.,    Ramdass V.~J.,  2008, The Astrophysical Journal,
  687, 39

\bibitem[\protect\citeauthoryear{{Diego}, {Protopapas}, {Sandvik} \&
  {Tegmark}}{{Diego} et~al.}{2005}]{diego2005a}
{Diego} J.~M.,  {Protopapas} P.,  {Sandvik} H.~B.,    {Tegmark} M.,  2005,
  \mnras, 360, 477

\bibitem[\protect\citeauthoryear{{Diego}, {Tegmark}, {Protopapas} \&
  {Sandvik}}{{Diego} et~al.}{2007}]{diego2007}
{Diego} J.~M.,  {Tegmark} M.,  {Protopapas} P.,    {Sandvik} H.~B.,  2007,
  \mnras, 375, 958

\bibitem[\protect\citeauthoryear{{El{\'{\i}}asd{\'o}ttir}, {Limousin},
  {Richard}, {Hjorth}, {Kneib}, {Natarajan}, {Pedersen}, {Jullo} \&
  {Paraficz}}{{El{\'{\i}}asd{\'o}ttir} et~al.}{2009}]{eliasdottir2007b}
{El{\'{\i}}asd{\'o}ttir} {\'A}.,  {Limousin} M.,  {Richard} J.,  {Hjorth} J.,
  {Kneib} J.-P.,  {Natarajan} P.,  {Pedersen} K.,  {Jullo} E.,    {Paraficz}
  D.,  2009, ArXiv e-prints

\bibitem[\protect\citeauthoryear{{Fu}, {Semboloni}, {Hoekstra}, {Kilbinger},
  {van Waerbeke}, {Tereno}, {Mellier}, {Heymans}, {Coupon}, {Benabed},
  {Benjamin}, {Bertin}, {Dor{\'e}}, {Hudson}, {Ilbert} \& {et~al.}}{{Fu}
  et~al.}{2008}]{fu2008}
{Fu} L.,  {Semboloni} E.,  {Hoekstra} H.,  {Kilbinger} M.,  {van Waerbeke} L.,
  {Tereno} I.,  {Mellier} Y.,  {Heymans} C.,  {Coupon} J.,  {Benabed} K.,
  {Benjamin} J.,  {Bertin} E.,  {Dor{\'e}} O.,  {Hudson} M.~J.,  {Ilbert} O.,
   {et~al.} 2008, \aap, 479, 9

\bibitem[\protect\citeauthoryear{{Gavazzi}, {Treu}, {Rhodes}, {Koopmans},
  {Bolton}, {Burles}, {Massey} \& {Moustakas}}{{Gavazzi}
  et~al.}{2007}]{gavazzi2007}
{Gavazzi} R.,  {Treu} T.,  {Rhodes} J.~D.,  {Koopmans} L.~V.~E.,  {Bolton}
  A.~S.,  {Burles} S.,  {Massey} R.~J.,    {Moustakas} L.~A.,  2007, \apj, 667,
  176

\bibitem[\protect\citeauthoryear{{Gentile}, {Tonini} \& {Salucci}}{{Gentile}
  et~al.}{2007}]{gentile2007}
{Gentile} G.,  {Tonini} C.,    {Salucci} P.,  2007, \aap, 467, 925

\bibitem[\protect\citeauthoryear{{Guzzo}, {Pierleoni}, {Meneux}, {Branchini},
  {Le F{\`e}vre}, {Marinoni}, {Garilli}, {Blaizot}, {De Lucia}, {Pollo},
  {McCracken}, {Bottini}, {Le Brun}, {Maccagni}, {Picat}, {Scaramella} \&
  {et~al.}}{{Guzzo} et~al.}{2008}]{guzzo2008}
{Guzzo} L.,  {Pierleoni} M.,  {Meneux} B.,  {Branchini} E.,  {Le F{\`e}vre} O.,
   {Marinoni} C.,  {Garilli} B.,  {Blaizot} J.,  {De Lucia} G.,  {Pollo} A.,
  {McCracken} H.~J.,  {Bottini} D.,  {Le Brun} V.,  {Maccagni} D.,  {Picat}
  J.~P.,  {Scaramella} R.,    {et~al.} 2008, \nat, 451, 541

\bibitem[\protect\citeauthoryear{{Halkola}, {Seitz} \& {Pannella}}{{Halkola}
  et~al.}{2006}]{halkola2006}
{Halkola} A.,  {Seitz} S.,    {Pannella} M.,  2006, \mnras, pp 1133--+

\bibitem[\protect\citeauthoryear{{Halkola}, {Seitz} \& {Pannella}}{{Halkola}
  et~al.}{2007}]{halkola2007}
{Halkola} A.,  {Seitz} S.,    {Pannella} M.,  2007, \apj, 656, 739

\bibitem[\protect\citeauthoryear{{Hu} \& {Dodelson}}{{Hu} \&
  {Dodelson}}{2002}]{hu2002}
{Hu} W.,  {Dodelson} S.,  2002, \araa, 40, 171

\bibitem[\protect\citeauthoryear{{Jee}, {Ford}, {Illingworth}, {White},
  {Broadhurst}, {Coe}, {Meurer}, {van der Wel}, {Ben{\'{\i}}tez}, {Blakeslee},
  {Bouwens}, {Bradley}, {Demarco}, {Homeier}, {Martel} \& {Mei}}{{Jee}
  et~al.}{2007}]{jee2007}
{Jee} M.~J.,  {Ford} H.~C.,  {Illingworth} G.~D.,  {White} R.~L.,  {Broadhurst}
  T.~J.,  {Coe} D.~A.,  {Meurer} G.~R.,  {van der Wel} A.,  {Ben{\'{\i}}tez}
  N.,  {Blakeslee} J.~P.,  {Bouwens} R.~J.,  {Bradley} L.~D.,  {Demarco} R.,
  {Homeier} N.~L.,  {Martel} A.~R.,    {Mei} S.,  2007, \apj, 661, 728

\bibitem[\protect\citeauthoryear{{Jullo}, {Kneib}, {Limousin},
  {El{\'{\i}}asd{\'o}ttir}, {Marshall} \& {Verdugo}}{{Jullo}
  et~al.}{2007}]{jullo2007}
{Jullo} E.,  {Kneib} J.-P.,  {Limousin} M.,  {El{\'{\i}}asd{\'o}ttir} {\'A}.,
  {Marshall} P.~J.,    {Verdugo} T.,  2007, New Journal of Physics, 9, 447

\bibitem[\protect\citeauthoryear{{Kassiola} \& {Kovner}}{{Kassiola} \&
  {Kovner}}{1993}]{kassiola1993}
{Kassiola} A.,  {Kovner} I.,  1993, \apj, 417, 450

\bibitem[\protect\citeauthoryear{{Kneib}, {Ellis}, {Smail}, {Couch} \&
  {Sharples}}{{Kneib} et~al.}{1996}]{kneib1996}
{Kneib} J.-P.,  {Ellis} R.~S.,  {Smail} I.,  {Couch} W.~J.,    {Sharples}
  R.~M.,  1996, \apj, 471, 643

\bibitem[\protect\citeauthoryear{{Kneib}, {Mellier}, {Fort} \&
  {Mathez}}{{Kneib} et~al.}{1993}]{kneib1993}
{Kneib} J.~P.,  {Mellier} Y.,  {Fort} B.,    {Mathez} G.,  1993, \aap, 273, 367

\bibitem[\protect\citeauthoryear{{Koopmans}, {Treu}, {Bolton}, {Burles} \&
  {Moustakas}}{{Koopmans} et~al.}{2006}]{koopmans2006}
{Koopmans} L.~V.~E.,  {Treu} T.,  {Bolton} A.~S.,  {Burles} S.,    {Moustakas}
  L.~A.,  2006, \apj, 649, 599

\bibitem[\protect\citeauthoryear{{Liesenborgs}, {de Rijcke}, {Dejonghe} \&
  {Bekaert}}{{Liesenborgs} et~al.}{2007}]{liesenborgs2007}
{Liesenborgs} J.,  {de Rijcke} S.,  {Dejonghe} H.,    {Bekaert} P.,  2007,
  \mnras, 380, 1729

\bibitem[\protect\citeauthoryear{{Liesenborgs}, {de Rijcke}, {Dejonghe} \&
  {Bekaert}}{{Liesenborgs} et~al.}{2008}]{liesenborgs2008a}
{Liesenborgs} J.,  {de Rijcke} S.,  {Dejonghe} H.,    {Bekaert} P.,  2008,
  \mnras, 386, 307

\bibitem[\protect\citeauthoryear{{Limousin}, {Kneib} \& {Natarajan}}{{Limousin}
  et~al.}{2005}]{limousin2005}
{Limousin} M.,  {Kneib} J.-P.,    {Natarajan} P.,  2005, \mnras, 356, 309

\bibitem[\protect\citeauthoryear{{Limousin}, {Richard}, {Jullo}, {Kneib},
  {Fort}, {Soucail}, {El{\'{\i}}asd{\'o}ttir}, {Natarajan}, {Ellis}, {Smail},
  {Czoske}, {Smith}, {Hudelot}, {Bardeau}, {Ebeling}, {Egami} \&
  {Knudsen}}{{Limousin} et~al.}{2007}]{limousin2007b}
{Limousin} M.,  {Richard} J.,  {Jullo} E.,  {Kneib} J.~P.,  {Fort} B.,
  {Soucail} G.,  {El{\'{\i}}asd{\'o}ttir} A.,  {Natarajan} P.,  {Ellis} R.~S.,
  {Smail} I.,  {Czoske} O.,  {Smith} G.~P.,  {Hudelot} P.,  {Bardeau} S.,
  {Ebeling} H.,  {Egami} E.,    {Knudsen} K.~K.,  2007, \apj, 668, 643

\bibitem[\protect\citeauthoryear{{Limousin}, {Richard}, {Kneib}, {Brink},
  {Pell{\'o}}, {Jullo}, {Tu}, {Sommer-Larsen}, {Egami}, {Micha{\l}owski},
  {Cabanac} \& {Stark}}{{Limousin} et~al.}{2008}]{limousin2008}
{Limousin} M.,  {Richard} J.,  {Kneib} J.-P.,  {Brink} H.,  {Pell{\'o}} R.,
  {Jullo} E.,  {Tu} H.,  {Sommer-Larsen} J.,  {Egami} E.,  {Micha{\l}owski}
  M.~J.,  {Cabanac} R.,    {Stark} D.~P.,  2008, \aap, 489, 23

\bibitem[\protect\citeauthoryear{{Mahdavi}, {Hoekstra}, {Babul}, {Balam} \&
  {Capak}}{{Mahdavi} et~al.}{2007}]{mahdavi2007}
{Mahdavi} A.,  {Hoekstra} H.,  {Babul} A.,  {Balam} D.~D.,    {Capak} P.~L.,
  2007, \apj, 668, 806

\bibitem[\protect\citeauthoryear{{Marshall}, {Hobson}, {Gull} \&
  {Bridle}}{{Marshall} et~al.}{2002}]{marshall2002}
{Marshall} P.~J.,  {Hobson} M.~P.,  {Gull} S.~F.,    {Bridle} S.~L.,  2002,
  \mnras, 335, 1037

\bibitem[\protect\citeauthoryear{{Massey}, {Rhodes}, {Ellis}, {Scoville},
  {Leauthaud}, {Finoguenov}, {Capak}, {Bacon}, {Aussel}, {Kneib}, {Koekemoer},
  {McCracken}, {Mobasher}, {Pires}, {Refregier}, {Sasaki}, {Starck} \&
  {Taniguchi}}{{Massey} et~al.}{2007}]{massey2007nat}
{Massey} R.,  {Rhodes} J.,  {Ellis} R.,  {Scoville} N.,  {Leauthaud} A.,
  {Finoguenov} A.,  {Capak} P.,  {Bacon} D.,  {Aussel} H.,  {Kneib} J.-P.,
  {Koekemoer} A.,  {McCracken} H.,  {Mobasher} B.,  {Pires} S.,  {Refregier}
  A.,  {Sasaki} S.,  {Starck} J.-L.,    {Taniguchi} 2007, \nat, 445, 286

\bibitem[\protect\citeauthoryear{{Meneghetti}, {Argazzi}, {Pace}, {Moscardini},
  {Dolag}, {Bartelmann}, {Li} \& {Oguri}}{{Meneghetti}
  et~al.}{2007}]{meneghetti2007}
{Meneghetti} M.,  {Argazzi} R.,  {Pace} F.,  {Moscardini} L.,  {Dolag} K.,
  {Bartelmann} M.,  {Li} G.,    {Oguri} M.,  2007, \aap, 461, 25

\bibitem[\protect\citeauthoryear{{Natarajan}, {De Lucia} \&
  {Springel}}{{Natarajan} et~al.}{2007}]{natarajan2007a}
{Natarajan} P.,  {De Lucia} G.,    {Springel} V.,  2007, \mnras, 376, 180

\bibitem[\protect\citeauthoryear{{Navarro}, {Frenk} \& {White}}{{Navarro}
  et~al.}{1997}]{navarro1997}
{Navarro} J.~F.,  {Frenk} C.~S.,    {White} S.~D.~M.,  1997, \apj, 490, 493

\bibitem[\protect\citeauthoryear{{Saha} \& {Williams}}{{Saha} \&
  {Williams}}{1997}]{saha1997}
{Saha} P.,  {Williams} L.~L.~R.,  1997, \mnras, 292, 148

\bibitem[\protect\citeauthoryear{{Salucci}}{{Salucci}}{2001}]{salucci2001}
{Salucci} P.,  2001, \mnras, 320, L1

\bibitem[\protect\citeauthoryear{{Sand}, {Treu}, {Ellis}, {Smith} \&
  {Kneib}}{{Sand} et~al.}{2008}]{sand2008}
{Sand} D.~J.,  {Treu} T.,  {Ellis} R.~S.,  {Smith} G.~P.,    {Kneib} J.-P.,
  2008, \apj, 674, 711

\bibitem[\protect\citeauthoryear{{Schneider}, {Ehlers} \& {Falco}}{{Schneider}
  et~al.}{1992}]{schneider1992}
{Schneider} P.,  {Ehlers} J.,    {Falco} E.~E.,  1992, {Gravitational Lenses}.
Gravitational Lenses, XIV, 560 pp.~112 figs..~Springer-Verlag Berlin Heidelberg
  New York.~ Also Astronomy and Astrophysics Library

\bibitem[\protect\citeauthoryear{{Seljak}, {Makarov}, {McDonald}, {Anderson},
  {Bahcall}, {Brinkmann}, {Burles}, {Cen}, {Doi}, {Gunn}, {Ivezi{\'c}}, {Kent},
  {Loveday}, {Lupton}, {Munn}, {Nichol} \& {et~al.}}{{Seljak}
  et~al.}{2005}]{seljak2005}
{Seljak} U.,  {Makarov} A.,  {McDonald} P.,  {Anderson} S.~F.,  {Bahcall}
  N.~A.,  {Brinkmann} J.,  {Burles} S.,  {Cen} R.,  {Doi} M.,  {Gunn} J.~E.,
  {Ivezi{\'c}} {\v Z}.,  {Kent} S.,  {Loveday} J.,  {Lupton} R.~H.,  {Munn}
  J.~A.,  {Nichol} R.~C.,    {et~al.} 2005, \prd, 71, 103515

\bibitem[\protect\citeauthoryear{{Simon} \& {Geha}}{{Simon} \&
  {Geha}}{2007}]{simon2007}
{Simon} J.~D.,  {Geha} M.,  2007, \apj, 670, 313

\bibitem[\protect\citeauthoryear{{Skilling}}{{Skilling}}{2004}]{skilling2004}
{Skilling} J.,  2004, {BayeSys and MassInf}

\bibitem[\protect\citeauthoryear{{Smith} \& {Taylor}}{{Smith} \&
  {Taylor}}{2008}]{smith2008}
{Smith} G.~P.,  {Taylor} J.~E.,  2008, \apjl, 682, L73

\bibitem[\protect\citeauthoryear{{Spergel} \& {Steinhardt}}{{Spergel} \&
  {Steinhardt}}{2000}]{spergel2000}
{Spergel} D.~N.,  {Steinhardt} P.~J.,  2000, Physical Review Letters, 84, 3760

\bibitem[\protect\citeauthoryear{{Spergel}, {Verde}, {Peiris}, {Komatsu},
  {Nolta}, {Bennett}, {Halpern}, {Hinshaw}, {Jarosik}, {Kogut}, {Limon},
  {Meyer}, {Page}, {Tucker}, {Weiland}, {Wollack} \& {Wright}}{{Spergel}
  et~al.}{2003}]{spergel2003}
{Spergel} D.~N.,  {Verde} L.,  {Peiris} H.~V.,  {Komatsu} E.,  {Nolta} M.~R.,
  {Bennett} C.~L.,  {Halpern} M.,  {Hinshaw} G.,  {Jarosik} N.,  {Kogut} A.,
  {Limon} M.,  {Meyer} S.~S.,  {Page} L.,  {Tucker} G.~S.,  {Weiland} J.~L.,
  {Wollack} E.,    {Wright} E.~L.,  2003, \apjs, 148, 175

\bibitem[\protect\citeauthoryear{{Suyu}, {Marshall}, {Hobson} \&
  {Blandford}}{{Suyu} et~al.}{2006}]{suyu2006}
{Suyu} S.~H.,  {Marshall} P.~J.,  {Hobson} M.~P.,    {Blandford} R.~D.,  2006,
  \mnras, 371, 983

\bibitem[\protect\citeauthoryear{{Valenzuela}, {Rhee}, {Klypin}, {Governato},
  {Stinson}, {Quinn} \& {Wadsley}}{{Valenzuela} et~al.}{2007}]{valenzuela2007}
{Valenzuela} O.,  {Rhee} G.,  {Klypin} A.,  {Governato} F.,  {Stinson} G.,
  {Quinn} T.,    {Wadsley} J.,  2007, \apj, 657, 773

\end{thebibliography}

\end{document}